\documentclass[12pt]{article}
\usepackage[british]{babel} 

\def\AnswerYes{y} 
\def\pdflatex{y} 
\def\ShowLabelsVersion{n} 
\def\ShowChangesVersion{n} 
\def\ShowAnnotationsVersion{n} 

\def\feynVersion{n} 
\usepackage[final]{graphicx} 

\ifx\pdflatex\AnswerYes
\usepackage[update,prepend]{epstopdf} 
\usepackage{grffile} 
\ifx\pdflatex\AnswerYes \usepackage{color} \else \usepackage[dvips]{color} \fi
\ifx\pdflatex\AnswerYes \usepackage{thumbpdf} %
\else \usepackage[dvips]{thumbpdf} \fi
\ifx\pdflatex\AnswerYes \usepackage[ final, breaklinks=true,
colorlinks=true, 
pdfborder={0 0 1}, 
citecolor={blue},linkcolor={blue},urlcolor={red}, citebordercolor={1 0
  0},linkbordercolor={0 0 1},urlbordercolor={1 0 0},
pdfpagemode=UseOutlines,
bookmarks=true,bookmarksopenlevel=4
]{hyperref} 
\else \usepackage[dvips, 
final, breaklinks=true, colorlinks=true, 
pdfborder={0 0 1}, 
citecolor={blue},linkcolor={blue},urlcolor={red}, citebordercolor={1 0
  0},linkbordercolor={0 0 1},urlbordercolor={1 0 0},
pdfpagemode=UseOutlines,
bookmarks=true,bookmarksopenlevel=4
]{hyperref} 
\fi
\usepackage[numbers,sort&compress]{natbib}

\usepackage{multirow} 
\usepackage{amssymb} \usepackage{amsmath}
\usepackage{xspace} 
\usepackage{dcolumn} 
\usepackage{bm} 

\usepackage{slashed}
\ifx\feynVersion\AnswerYes
\usepackage{feynmp} 
\fi

\usepackage{pbox}

\ifx\ShowLabelsVersion\AnswerYes
\usepackage[color]{showkeys}
\definecolor{refkey}{gray}{.5} 
\definecolor{labelkey}{gray}{.5} 

\fi

\ifx\ShowAnnotationsVersion\AnswerYes
\newcommand{\comment}[1]{{\scriptsize\sffamily\bfseries{#1}}}
\newcommand{\margin}[1]{\marginpar{\scriptsize\sffamily\bfseries{#1}}}
 \newcommand{\drafty}{\textbf{Draft version \today}
  \hfill} \else \newcommand{\comment}[1]{} \newcommand{\margin}[1]{}
\newcommand{\drafty}{} \fi

\ifx\ShowChangesVersion\AnswerYes
 \usepackage{ulem}

\newcommand{\delete}[1]{\sout{#1}} 
\renewcommand{\emph}[1]{\textit{#1}} 
\else  
 
 \newcommand{\sout}[1]{} \newcommand{\xout}[1]{}
 
\newcommand{\delete}[1]{} \fi

\newcommand{\absatz}{\vspace{2ex}\noindent}

\newcommand{\dis}{\displaystyle} 

\newcommand{\non}{\nonumber} 
 \newcommand{\hqq}{\hspace{1em}}
\newcommand{\hqqq}{\hspace{2em}} 
\newcommand{\phm}{\phantom{-}} 

\newcommand{\half}{\frac{1}{2}} 
 \newcommand{\dd}{\mathrm{d}}

\def\lsim{\mathrel{\rlap{\lower4pt\hbox{\hskip1pt$\sim$}}
    \raise1pt\hbox{$<$}}} 
\def\gsim{\mathrel {\rlap{\lower4pt\hbox{\hskip1pt$\sim$}}
    \raise1pt\hbox{$>$}}} 

\newcommand{\mpi}{\ensuremath{m_\pi}} \newcommand{\fpi}{\ensuremath{f_\pi}}
\newcommand{\mpiphys}{\ensuremath{m_\pi^\text{phys}}}
\newcommand{\MeV}{\ensuremath{\mathrm{MeV}}}
\newcommand{\fm}{\ensuremath{\mathrm{fm}}}
\newcommand{\ChiEFT}{$\chi$EFT\xspace}

\newcommand{\NXLO}[1]{N\ensuremath{{}^{#1}}LO\xspace}

\newcommand{\HIGS}{HI$\gamma$S\xspace}

\newcommand{\alphae}{\ensuremath{\alpha_{E1}}}
\newcommand{\betam}{\ensuremath{\beta_{M1}}}
\newcommand{\gammaee}{\ensuremath{\gamma_{E1E1}}}
\newcommand{\gammamm}{\ensuremath{\gamma_{M1M1}}}
\newcommand{\gammaem}{\ensuremath{\gamma_{E1M2}}}
\newcommand{\gammame}{\ensuremath{\gamma_{M1E2}}}
\newcommand{\gammazero}{\ensuremath{\gamma_{0}}}
\newcommand{\gammapi}{\ensuremath{\gamma_{\pi}}}
\newcommand{\alphaep}{\ensuremath{\alpha_{E1}^{(\mathrm{p})}}}
\newcommand{\betamp}{\ensuremath{\beta_{M1}^{(\mathrm{p})}}}
\newcommand{\gammaeep}{\ensuremath{\gamma_{E1E1}^{(\mathrm{p})}}}
\newcommand{\gammammp}{\ensuremath{\gamma_{M1M1}^{(\mathrm{p})}}}
\newcommand{\gammaemp}{\ensuremath{\gamma_{E1M2}^{(\mathrm{p})}}}
\newcommand{\gammamep}{\ensuremath{\gamma_{M1E2}^{(\mathrm{p})}}}
\newcommand{\gammazerop}{\ensuremath{\gamma_{0}^{(\mathrm{p})}}}
\newcommand{\gammapip}{\ensuremath{\gamma_{\pi}^{(\mathrm{p})}}}
\newcommand{\alphaen}{\ensuremath{\alpha_{E1}^{(\mathrm{n})}}}
\newcommand{\betamn}{\ensuremath{\beta_{M1}^{(\mathrm{n})}}}
\newcommand{\gammaeen}{\ensuremath{\gamma_{E1E1}^{(\mathrm{n})}}}
\newcommand{\gammammn}{\ensuremath{\gamma_{M1M1}^{(\mathrm{n})}}}
\newcommand{\gammaemn}{\ensuremath{\gamma_{E1M2}^{(\mathrm{n})}}}
\newcommand{\gammamen}{\ensuremath{\gamma_{M1E2}^{(\mathrm{n})}}}
\newcommand{\gammazeron}{\ensuremath{\gamma_{0}^{(\mathrm{n})}}}
\newcommand{\gammapin}{\ensuremath{\gamma_{\pi}^{(\mathrm{n})}}}
\newcommand{\alphaes}{\ensuremath{\alpha_{E1}^{(\mathrm{s})}}}
\newcommand{\betams}{\ensuremath{\beta_{M1}^{(\mathrm{s})}}}

\newcommand{\alphaev}{\ensuremath{\alpha_{E1}^{(\mathrm{v})}}}
\newcommand{\betamv}{\ensuremath{\beta_{M1}^{(\mathrm{v})}}}

\newcommand{\kappas}{\ensuremath{\kappa^{(\mathrm{s})}}}
\newcommand{\kappav}{\ensuremath{\kappa^{(\mathrm{v})}}}
\newcommand{\kappap}{\ensuremath{\kappa^{(\mathrm{p})}}}
\newcommand{\kappan}{\ensuremath{\kappa^{(\mathrm{n})}}}

\newcommand{\MN}{\ensuremath{M_\mathrm{N}}} 
\newcommand{\MDelta}{\ensuremath{M_\Delta}} 
\newcommand{\DeltaM}{\ensuremath{\Delta_{\scriptscriptstyle
      M}}} 

\newcommand{\ga}{g_{\scriptscriptstyle A}}
\newcommand{\gpiNN}{g_{\pi{\scriptscriptstyle\text{NN}}}}
\newcommand{\gpiNDelta}{g_{\pi{\scriptscriptstyle\text{N}\Delta}}}
\newcommand{\lambdachi}{\Lambda_\chi}  \newcommand{\alphaEM}{\alpha_{\scriptscriptstyle\text{EM}}}

\newcommand{\pr}{{\rm pr}}

 \newcommand{\calL}{\mathcal{L}}
 
\newcommand{\calO}{\mathcal{O}}

\newcommand{\mytitle}[1]{\begin{center}\LARGE{\textbf{#1}}\end{center}}
\newcommand{\myauthor}[1]{\textbf{#1}} \newcommand{\myaddress}[1]{\textit{#1}}
\newcommand{\mypreprint}[1]{\begin{flushright}#1\end{flushright}}

\begin{document}

%

\begin{titlepage}
  \setcounter{page}{0} \mypreprint{
    \drafty
    6th November 2015 \\
    Final version 18 May 2016, accepted by
    Eur.~Phys.~J.~\textbf{A}.
  }
  
  \mytitle{Nucleon Polarisabilities At and Beyond Physical Pion Masses}
  

\begin{center}
  \myauthor{Harald W.\ Grie\3hammer$^{a}$}\footnote{Email:
    hgrie@gwu.edu}, 
  \myauthor{Judith A.~McGovern$^{b}$}\footnote{Email:
    judith.mcgovern@manchester.ac.uk} 
  \emph{and} 
  \myauthor{Daniel R.~Phillips$^{c}$}\footnote{Email: phillips@phy.ohiou.edu}
  
  \vspace*{0.2cm}
  
  \myaddress{$^a$ Institute for Nuclear Studies, Department of Physics, \\The
    George Washington University, Washington DC 20052, USA}
  \\[1ex]
  \myaddress{$^b$ School of Physics and Astronomy, The University of
    Manchester, \\Manchester M13 9PL, UK}
  \\[1ex]
  \myaddress{$^c$ Department of Physics and Astronomy and Institute of Nuclear
    and Particle Physics, Ohio University, Athens, Ohio 45701, USA}

  \vspace*{0.2cm}

\end{center}


\begin{abstract} 
  We examine the results of Chiral Effective Field Theory (\ChiEFT) for the
  scalar- and spin-dipole polarisabilities of the proton and neutron, both for
  the physical pion mass and as a function of $\mpi$. This provides chiral
  extrapolations for lattice-QCD polarisability computations. We include both
  the leading and sub-leading effects of the nucleon's pion cloud, as well as
  the leading ones of the $\Delta(1232)$ resonance and its pion cloud. The
  analytic results are complete at \NXLO{2} in the $\delta$-counting for pion
  masses close to the physical value, and at leading order for pion masses
  similar to the Delta-nucleon mass splitting.  In order to quantify the
  truncation error of our predictions and fits as $68$\% degree-of-belief
  intervals, we use a Bayesian procedure recently adapted to EFT
  expansions. At the physical point, our predictions for the spin
  polarisabilities are, within respective errors, in good agreement with
  alternative extractions using experiments and dispersion-relation theory. At
  larger pion masses we find that the chiral expansion of all polarisabilities
  becomes intrinsically unreliable as $\mpi$ approaches about $300\;\MeV$---as
  has already been seen in other observables. \ChiEFT also predicts a
  substantial isospin splitting above the physical point for both the electric
  and magnetic scalar polarisabilities; and we speculate on the impact this
  has on the stability of nucleons. Our results agree very well with emerging
  lattice computations in the realm where \ChiEFT converges. Curiously, for
  the central values of some of our predictions, this agreement persists to
  much higher pion masses.  We speculate on whether this might be more than a
  fortuitous coincidence.
\end{abstract}
\noindent
\begin{tabular}{rl}
  Suggested Keywords: &\begin{minipage}[t]{10.7cm} Effective Field Theory,
    lattice QCD, chiral extrapolation, proton, neutron and nucleon
    polarisabilities, spin polarisabilities, Chiral Perturbation Theory,
    $\Delta(1232)$ resonance, Bayesian statistics, uncertainty/error estimates. 
  \end{minipage}
\end{tabular}


\end{titlepage}

\setcounter{footnote}{0}

\newpage


\section{Introduction}
\setcounter{equation}{0}
\label{sec:introduction}

The polarisabilities of a composite system are among its most basic
properties; see e.g.~\cite{Griesshammer:2012we} for a recent review. At a
classical level, they reflect how much freedom charged constituents have to
rearrange under the application of external electromagnetic fields, while in
quantum mechanics they indicate how easily electromagnetic interactions induce
transitions to low-lying excited states. They therefore encode information
about the symmetries and strengths of constituents' interactions with each
other and with the photon. As well as the usual electric ($\alphae$) and
magnetic ($\betam$) polarisabilities, a spin-half object like the nucleon has
four ``spin-polarisabilities" ($\gamma_i$). These are less obvious in their
effects but encode the spin-dependent response and can, for instance, be
related to effects analogous to birefringence and Faraday rotation for
long-wavelength electromagnetic radiation. In the nucleon, the lightest
relevant excitation involves the creation of a virtual charged pion. This
mechanism is expected to dominate the electric polarisability and contribute
significantly to others, too. The exploration of nucleon polarisabilities was
therefore a natural early application of Chiral Perturbation Theory in the
baryonic sector~\cite{Jenkins:1991ne,Bernard:1991rq,Bernard:1995dp} which
predicts the behaviour of each polarisability as it diverges in the chiral
limit $\mpi\to0$~\cite{Bernard:1991rq}. On the other hand, in the real world
the excitation energy of the $\Delta(1232)$ resonance, $\DeltaM \equiv
\MDelta-\MN$, is about $300\;\MeV$, and thus not very much larger than the
physical pion mass. Furthermore, the strong magnetic N$\Delta$ dipole
transition should give a large paramagnetic contribution to the magnetic
polarisability.

The inclusion of the Delta as an explicit degree of freedom in Chiral
Effective Field Theory~\cite{Jenkins:1991ne, Butler:1992ci, Hemmert:1996xg,
  Hemmert:1997ye} enables quantitative predictions to be made for Compton
scattering~\cite{Pascalutsa:2002pi, Hildebrandt:2003fm}. This EFT has recently
been used in the most accurate extant determinations of the electric and
magnetic polarisabilities of the proton and neutron from Compton scattering
data~\cite{Griesshammer:2012we, McGovern:2012ew, Myers:2014ace}.  This
progress in the theory of polarisabilities is coupled to an upsurge of
interest in new experiments that are devoted to obtaining or refining our
knowledge of all the polarisabilities, electric, magnetic and spin, of both
the proton and neutron~\cite{Weller:2009zz, HIGSPAC, Downie:2011mm,
  Huber:2015uza}, with results from MAXlab~\cite{Myers:2014ace, Myers:2015aba}
and MAMI~\cite{Martel:2014pba} published within the last year.

The calculation of nucleon polarisabilities directly from the QCD action is
also an aim of lattice QCD. The need to incorporate electromagnetic fields in
the computation creates challenges, which means that this is a fairly new
endeavour, but several groups now have published
results~\cite{Chang:2015qxa,Lujan:2014qga,Detmold:2010ts,Primer:2013pva,
  Hall:2013dva,Engelhardt:2011qq,Engelhardt:2007ub,Engelhardt:2010tm,
  Engelhardt:2015,Freeman:2014kka}.  Since all are at pion masses
substantially above the physical pion mass, the question of how to extrapolate
to the real world is of pressing interest, and can be addressed within
\ChiEFT. Our analysis provides a bridge between data and lattice QCD, where a
direct computation of Compton scattering would be highly nontrivial.

Polarisabilities are therefore fundamental characteristics of hadrons, and
benchmarks for our understanding of hadronic structure; a summary of their
importance and best ways to access them was also provided by a number of
theorists in Ref.~\cite{Griesshammer:2014xla}. Furthermore, their values have
other implications, some examples of which we now discuss. 
First, the Cottingham Sum rule relates the doubly-virtual forward Compton
scattering amplitude, and hence the proton-neutron difference in $\betam$, to
the proton-neutron electromagnetic mass difference~\cite{WalkerLoud:2012bg,
  WalkerLoud:2012en, Erben:2014hza,Thomas:2014dxa,Gasser:2015dwa}.  The
relation between the mass difference and the polarisabilities proceeds via a
low-energy theorem for the subtraction function in the Cottingham formula at
vanishing momentum, which is related to
$\betam^{(\text{p-n})}$~\cite{WalkerLoud:2012bg,Gasser:2015dwa}. When one uses
present knowledge on $\betam^{(\text{p-n})}$ as input and models the
subtraction function along the lines suggested in
Refs.~\cite{WalkerLoud:2012bg, WalkerLoud:2012en, Erben:2014hza}, the
uncertainty in the polarisability contributes sizeably to the uncertainty in
the mass difference.  Conversely, assuming knowledge about the electromagnetic
part of the mass difference provides a constraint on the
polarisabilities~\cite{Thomas:2014dxa}.
Either scenario tests our understanding of the subtle interplay between
electromagnetic and strong interactions in a fundamental observable.
Second, the magnetic polarisability, $\betam$, is also crucial for the
two-photon-exchange contribution to the Lamb shift in muonic
hydrogen~\cite{Pachucki, Carlson:2011dz, Pohl:2013yb}, the least-known
ingredient of the ``proton-radius puzzle''.

The aim of this paper is thus two-fold. Firstly, we will present the analytic
expressions and numerical results for all static dipole polarisabilities as
they enter in the Compton amplitudes used in the recent proton and neutron
analyses~\cite{McGovern:2012ew, Myers:2014ace}.  There is considerable
evidence that the extraction of $\alphaep$ and $\betamp$ from unpolarised
Compton scattering is robust against variations in the spin
polarisabilities~\cite{Griesshammer:2012we, McGovern:2012ew,
  Lensky:2014efa}. This means that polarisation observables are the best place
to determine these latter quantities~\cite{observables}, and programs at MAMI
and HI$\gamma$S are engaged in that pursuit~\cite{Weller:2009zz, HIGSPAC,
  Downie:2011mm, Huber:2015uza}. Secondly, we use our expressions to predict
the running of the polarisabilities with the pion mass, the better to compare
with lattice computations at numerically less costly, heavier, pion masses.

In both these contexts, we pay particular attention to the uncertainties of
our predictions and extractions, which are of two types. The impact of
statistical errors on data on \ChiEFT parameters can ultimately be reduced by
experimental and non-EFT-related efforts. However, there is also a
``truncation error'' which is intrinsic to an EFT, and it is that we focus on
in this paper. Because \ChiEFT gives a perturbative series for all
polarisabilities, this truncation error accounts for the fact that we only
have computed up to a finite order in the EFT
expansion~\cite{Cacciari:2011ze,Furnstahl:2015rha}.  Without its proper
appraisal---and that of any other uncertainties entering the \ChiEFT result---the significance of any agreement or discrepancy between theory and
experiment cannot be assessed~\cite{Furnstahl:2014xsa}.

On a technical note we mention here that although polarisabilities are
frequency-depen\-dent functions (see e.g.~\cite{Hildebrandt:2003fm,
  Babusci:1998ww, Griesshammer:2012we}), this paper is concerned with the
static values, that is the limit as $\omega\to0$. We will report these in the
canonical units of $10^{-4}~{\rm fm}^3$ for the scalar polarisabilities, and
$10^{-4}~{\rm fm}^4$ for the spin ones. Preliminary findings were reported in
Ref.~\cite{talkMAMI} and provided for inclusion in Refs.~\cite{Martel:2014pba,
  Lujan:2014qga}.

\absatz The presentation is organised as follows. In
Sect.~\ref{sec:formalism}, we define the chiral power counting in the regimes
relevant for lattice computations and summarise the analytic results for the
scalar and spin polarisabilities of the proton and neutron. After presenting
their values and uncertainties for physical pion masses in
Sect.~\ref{sec:values}, we detail our procedure to assign Bayesian
degree-of-belief intervals (Sects.~\ref{sec:theoryerrors}
and~\ref{sec:Finding-Theory-Uncertainties}) and discuss convergence
checks. Section~\ref{sec:extrapolations} extends this procedure to pion masses
above the physical point, provides predictions with error bars in
Fig.~\ref{fig:allpols}, and concludes with speculations on the relationship of
our findings to the proton-neutron mass splitting and anthropic arguments. We
then compare with available lattice computations in Sect.~\ref{sec:lattice},
and add Conclusions and an Appendix.

\section{\texorpdfstring{\ChiEFT}{Chiral EFT} with Dynamical
  \texorpdfstring{$\Delta(1232)$}{Delta(1232)} for Polarisabilities}
\setcounter{equation}{0}
\label{sec:formalism}

\subsection{Chiral Regimes and Power Counting}
\label{sec:regimes}

Compton scattering on nucleons in \ChiEFT has been reviewed in
Refs.~\cite{Griesshammer:2012we, McGovern:2012ew}, to which we refer the
reader for notation and the relevant parts of the chiral Lagrangian. Here, we
briefly discuss the power counting, which is crucial for our considerations,
and sketch the results.

Recall that Compton scattering exhibits three typical low-energy scales in
\ChiEFT with a dynamical Delta: the pion mass $\mpi$ as the typical chiral
scale; the Delta-nucleon mass splitting $\DeltaM \approx 300\;\MeV$; and the
photon energy $\omega$. Each provides a small, dimensionless expansion
parameter when measured in units of a natural ``high'' scale
$\Lambda_\chi\gg\DeltaM,\mpi,\omega$ at which the theory is to be expected to
break down because new degrees of freedom enter. For static scalar
polarisabilities, one considers the part of the amplitude which is quadratic
in $\omega$ as $\omega\to0$, and for spin polarisabilities the one cubic in
$\omega$. That leaves two parameters:
\begin{equation}
  P(\mpi) \equiv \frac{\mpi}{\Lambda_\chi} \qquad 
  \epsilon \equiv \frac{\MDelta-\MN}{\Lambda_\chi}\approx0.4\;\;,
  \label{eq:expparams}
\end{equation} 
where for simplicity we take one common breakdown scale $\Lambda_\chi\approx
650\;\MeV$ for both expansions, consistent with the masses of the $\omega$ and
$\rho$ as the next-lightest exchange mesons; we also count
$\MN\sim\Lambda_\chi$. This scale is only weakly dependent on $\mpi$, and we
will treat it as constant.

The fact that these two expansion parameters have a very different functional
dependence on $\mpi$ has important consequences for chiral extrapolations. The
Delta-nucleon mass splitting depends only weakly on the pion mass, and hence
$\epsilon$ is independent of $\mpi$ at the order to which we work. By
definition, though, the chiral parameter $P(\mpi)$ does change significantly
with $\mpi$.  We therefore identify three regimes relevant in contemporary
lattice computations, based on the relative size of $P$ and $\epsilon$. We
stress that regimes are not clearly separated, but transition from one regime
to the next is gradual.

In \textbf{regime (i)}, around the physical pion mass, $\mpi\approx\mpiphys$,
we follow Pascalutsa and Phillips~\cite{Pascalutsa:2002pi} and exploit a
numerical coincidence to define a single expansion parameter $\delta$:
\begin{equation}
  \label{eq:deltacountingi}
  \mbox{regime (i): }       
  \delta\approx\epsilon\approx\sqrt{P(\mpiphys)}\approx0.4\;\;.
\end{equation}
This is, of course, the regime relevant for the analysis of real-world Compton
scattering data, and hence this power counting determines the contributions
which were included in Refs.~\cite{Griesshammer:2012we, McGovern:2012ew} which
should be consulted for more details.

As the pion mass increases, we move into \textbf{regime (ii)},
$\mpi\approx\DeltaM\approx 300\;\MeV$. The two expansion parameters are now
numerically of comparable size, $P(\mpi)\approx\epsilon$, but their $\mpi$
dependence is still different.  It is then appropriate to identify
\begin{equation}
  \label{eq:deltacountingii}
  \mbox{regime (ii): }       
  \epsilon\approx P(\mpi\approx\DeltaM)\approx 0.4
\end{equation}
as the sole expansion parameter~\cite{Jenkins:1991ne, Butler:1992ci,
  Hemmert:1996xg, Hemmert:1997ye}.
  
Finally in \textbf{regime (iii)}, $\mpi\to\Lambda_\chi$, \ChiEFT becomes
inapplicable because the chiral expansion does not converge. A chiral
extrapolation of any observable can be expected to hold qualitatively at
best. In Sect.~\ref{sec:lattice}, we will see that \ChiEFT's polarisabilities
agree in the main with extant lattice results at such pion masses, but why
this should be so is unclear. The corresponding uncertainties are certainly
impossible to quantify with present techniques.

Note that we will not discuss the power counting for very small pion masses,
$\mpi \ll \MDelta - \MN$.  This regime near the chiral limit is fascinating,
but it will be some time before lattice computations explore it.

\subsection{Dipole Polarisabilities in Regime (i)}
\label{sec:pols}

Here we bring together the relevant expressions for the various contributions
to the polarisabilities: $\pi$N loops (calculated to subleading order),
$\pi\Delta$ loops, Delta pole diagrams, and low-energy constants from the
fourth-order $\pi$N Lagrangian.  The expressions for the dipole
polarisabilities are mostly published, but they are scattered in the
literature. The numerical values of all variables are listed in
Refs.~\cite{Griesshammer:2012we, McGovern:2012ew}\footnote{These contain a
  merely typographical error for $\kappas=-0.12$.}.

When we cite the expressions for all polarisabilities, we are excluding any
``non-structure" effects which would persist for a point-like nucleon with an
anomalous magnetic moment.  For the spin polarisabilities we also exclude the
contribution of the $\pi^0$ pole~\cite{Griesshammer:2012we}. These definitions
are standard in the literature, except that the backward spin polarisability
$\gammapi$ is often given without subtraction of the pion pole contribution of
$\mp45.9$ for the proton/neutron.  In terms of the multipole polarisabilities,
the forward and backward spin polarisabilities are defined as
\begin{equation}
  \label{eq:fbspin}
  \gammazero:=-\gammaee-\gammamm-\gammaem-\gammame\;\;,\;\;
  \gammapi:=-\gammaee+\gammamm-\gammaem+\gammame\;\;.
\end{equation}
We work in a heavy-baryon framework, except for the Delta pole, whose case is
explained shortly.  In $\delta$ counting, in regime (i), the leading
contribution to the Compton scattering amplitudes is the Thomson term which is
$\calO (e^2)$, followed by leading $\pi$N loops which are $\calO
(e^2\delta^2)$, then diagrams with a single Delta propagator which are $\calO
(e^2\delta^3)$, and finally subleading $\pi$N loops and LECs (counter terms)
at $\calO (e^2\delta^4)$.  In two cases, our expressions also contain terms
which are higher-order in $\delta$ counting. First, we do not expand the
$\pi\Delta$ loop expressions in powers of $\mpi/\DeltaM\approx\delta$; this
has the advantage that the expression remains valid as we move towards regime
(ii) $\mpi\sim \DeltaM$.  However, we do omit higher-order graphs that are
suppressed by $\DeltaM/\Lambda_\chi$ relative to the leading ones.  The other
exception is that we use a covariant Delta propagator for its pole graphs.
Since no loops or renormalisation are involved, this is simply a convenient
way of accounting for some kinematic higher-order effects, including the
electric $\gamma$N$\Delta$ coupling, which are relevant in the regime
$\omega\sim\DeltaM$.  For our present purposes these are not necessary, but as
they are small at the physical point and do not affect the running with
$\mpi$, we retain them for consistency with our previous work
\cite{McGovern:2012ew}.

Before presenting the formulae, we note a subtlety which arises when one
counts polarisabilities rather than amplitudes.  The electric and magnetic
polarisabilities $\alphae$ and $\betam$ follow the power-counting outlined
above; since in the amplitudes they multiply two powers of
$\omega\sim\mpi\sim\delta^2$, the contributions are $\calO
(e^2\delta^{-2})\sim \mpi^{-1}$ from the leading $\pi$N pieces, $\calO
(e^2\delta^{-1})\sim \DeltaM^{-1}$ from Delta contributions and $\calO
(e^2\delta^0)$ from subleading $\pi$N pieces, generating $\ln\mpi$ as well as
$\mpi$-independent contributions. On the other hand, the spin polarisabilities
$\gamma_i$ multiply three powers of $\omega\sim\mpi$ and hence start at $\calO
(e^2\delta^{-4})\sim \mpi^{-2}$. However, there are no contributions at $\calO
(e^2\delta^{-3})\sim (\mpi\DeltaM)^{-1}$ since the Delta-nucleon mass
difference acts as an infrared cut-off, forbidding Delta contributions to
diverge in the chiral limit. Instead, the Delta contributions start at $\calO
(e^2\delta^{-2})\sim \DeltaM^{-2}$, and these, together with the subleading
pion loops which are $\calO (e^2\delta^{-2})\sim \mpi^{-1}$, form the next
non-zero contribution to the (isoscalar) spin polarisabilities. This has
consequences for our error estimates, which are more reliable for $\alphae$
and $\betam$ where we have three nonzero terms in the $\delta$ expansion
series, than they are for the $\gamma_i$ where there are only two. In either
case, the last contribution calculated is of order $\delta^2$ relative to
leading.

\subsubsection{\texorpdfstring{$\pi$}{pi}N Loops}
\label{sec:Npiloops}

\begin{figure}[!b]
  \begin{center}
    \includegraphics[width=0.72\textwidth]{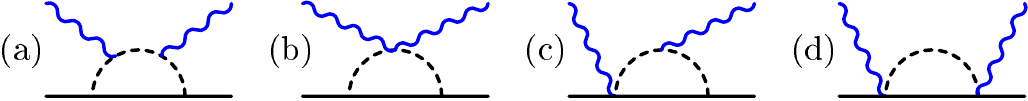}
    \caption{(Colour online) Leading contributions to the polarisabilities
      from the pion cloud around a nucleon in \ChiEFT. Interactions without
      symbol from $\calL_{\pi\mathrm{N}}^{(1)}$~\cite{Bernard:1995dp}.
      Permuted and crossed diagrams not displayed.}
    \label{fig:LOpiN}
  \end{center}
\end{figure}

The leading-order (LO) contributions from the pion cloud around the nucleon,
Fig.~\ref{fig:LOpiN}, were first calculated by Bernard, Kaiser and
Mei\3ner~\cite{Bernard:1991rq, Bernard:1995dp}:
\begin{align}
  &\alphae^{\pi\text{N, LO}} = 10 \betam^{\pi\text{N, LO}} =
  \frac{5\alphaEM\ga^2}{96\pi\fpi^2\mpi}\label{eq:abpiN-LO}\\
  &\gammaee^{\pi\text{N, LO}}=5\gammamm^{\pi\text{N, LO}}=
  -5\gammame^{\pi\text{N, LO}}=-5\gammaem^{\pi\text{N, LO}}=
  -\frac{5\alphaEM\ga^2}{96\pi^2\fpi^2\mpi^2}\;\;. \label{eq:gpiN-LO}
\end{align}
As motivated above, they diverge in the chiral limit and are indeed
$\calO(e^2\mpi^{-1}\sim e^2\delta^{-2})$ for the scalar polarisabilities, and
$\calO(e^2\mpi^{-2}\sim e^2\delta^{-4})$ for the spin ones.

For future reference we note that the $\pi^0$-pole contributes
\begin{equation}
  \gammaee^{\pi^0}=-\gammamm^{\pi^0}= \gammaem^{\pi^0}=-\gammame^{\pi^0}=\tau_3\;\frac{e^2\gpiNN}{16\pi^3\fpi\MN m_{\pi^0}^2}\;,
  \label{eq:pi-pole}
\end{equation}
where $\tau_3$ is the third Pauli matrix in isospin space.  This has the
numerical value of $11.5$ for the proton at the physical pion mass (with $\gpiNN^2/(4\pi)=13.64$~\cite{Rentmeester:1999vw}).
 
\begin{figure}[!b]
  \begin{center}
    \includegraphics[width=0.82\textwidth]{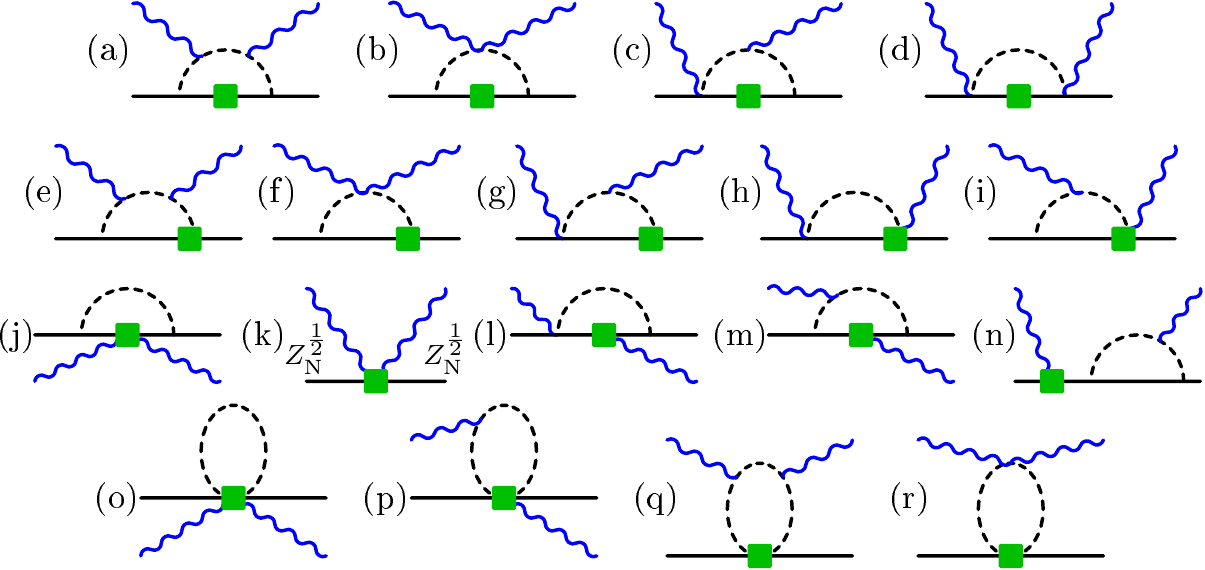}
    \caption{(Colour online) Subleading contributions to the polarisabilities
      from the pion cloud around a nucleon in \ChiEFT. Notation as in
      Fig.~\ref{fig:LOpiN}; square: vertex from
      $\calL_{\pi\mathrm{N}}^{(2)}$~\cite{Bernard:1995dp}. Permuted and
      crossed diagrams not displayed.}
    \label{fig:corrpiN}
  \end{center}
\end{figure}

The first chiral corrections, shown in Fig.~\ref{fig:corrpiN}, were found by
Bernard et al.~\cite{bksm93} for the scalar polarisabilities (logarithmic in
$\mpi$) and by Kumar, McGovern and Birse~\cite{VijayaKumar:2000pv} for the
spin ones (one inverse power of $\mpi$):
\begin{align}
  \alphae^{\pi\text{N, corr}} = & \frac{\alphaEM}{24\pi^2\fpi^2}\left[
    \left(\frac{2(3+\tau_3)\ga^2}{\MN}-c_2\right)\ln\frac{\mpi}{\mpiphys}+
    \left(\frac{(27+8\tau_3)\ga^2}{4\MN} -(2c_1+\frac{c_2}{2}-c_3)\right)
  \right]\non\\
  \betam^{\pi\text{N, corr}} = & \frac{\alphaEM}{24\pi^2\fpi^2}\left[
    \left(\frac{3(2+(1+\kappas)\tau_3)\ga^2}{\MN}-c_2\right)
    \ln\frac{\mpi}{\mpiphys}\right.
  \label{eq:abpiN-corr}\\
  &\left.\hspace{10ex}+\left(\frac{(13+6(1+\kappas)\tau_3)\ga^2}{4\MN}+
      (2c_1-\frac{c_2}{2}-c_3)\right)\right]\non
  \\
  \label{eq:gpiN-corr}
  \begin{split}
    \gammaee^{\pi\text{N, corr}} = & \frac{\alphaEM\ga^2}{384\pi\fpi^2\MN\mpi}
    11(2+\tau_3) \\
    \gammamm^{\pi\text{N, corr}} = & \frac{\alphaEM\ga^2}{384\pi\fpi^2\MN\mpi}
    \left(15+4\kappav+4(1+\kappas)\tau_3\right)\\
    \gammaem^{\pi\text{N, corr}} = & \frac{\alphaEM\ga^2}{384\pi\fpi^2\MN\mpi}
    (-6-\tau_3)\\
    \gammame^{\pi\text{N, corr}} = & \frac{\alphaEM\ga^2}{384\pi\fpi^2\MN\mpi}
    \left(-1+2\kappav-2(1+\kappas)\tau_3\right)\;\;,
  \end{split}
\end{align}
where $\kappas=\kappap+\kappan=-0.12$ and $\kappav=\kappap-\kappan=3.71$ are
the anomalous magnetic moments of the nucleon, and $c_{1,2,3}$ low-energy
constants from the next-to-leading order (NLO) $\pi$N Lagrangian, determined,
e.g.~from $\pi$N scattering.  We set the renormalisation scale in the chiral
logarithms to be $\mpiphys$.

To the order we work, these are the \emph{only} $\mpi$-dependent contributions
which contain an isovector component and hence differentiate between proton
and neutron polarisabilities. For $\alphae$ and $\betam$, the chiral logarithm
provides a parameter-free and rather strong $\mpi$-dependence in the
difference---besides an $\mpi$-independent offset. The pion-mass dependence
of the proton-neutron split in the spin polarisabilities is stronger, scaling
with $\mpi^{-1}$, but will turn out to be considerably smaller than the
theoretical uncertainties of our predictions. See discussions in
Sects.~\ref{sec:results}, \ref{sec:isovector} and~\ref{sec:lattice}.

\subsubsection{Low-Energy Coefficients}
\label{sec:CTs}

In addition to the loops, there are contributions to polarisabilities directly
from the $\mpi$- and $\DeltaM$-independent low-energy coefficients (LECs)
multiplying operators in the Lagrangian.  Their finite parts subsume physics
which is unrelated to the pion cloud or to the Delta, generically:
\begin{equation}
  \label{eq:scalarLEC}
  \xi^{\text{LEC}}=\xi^{\text{(s)LEC}}+\tau_3\, \xi^{\text{(v)LEC}}
\end{equation}
for the isoscalar and isovector LECs of any polarisability
$\xi\in\{\alphae,\betam,\gamma_i\}$.

Of these, only $\alphae^{\text{LEC}}$ and $\betam^{\text{LEC}}$ need to be
included as counter terms at the same order as the $\pi$N corrections to
$\alphae$ and $\betam$ of eqs.~\eqref{eq:abpiN-corr}; they absorb the
renormalisation-point dependence of the chiral logarithms induced by the
divergent loops of Fig.~\ref{fig:corrpiN}.

In Ref.~\cite{McGovern:2012ew}, we determined the proton values by fitting to
a statistically consistent proton Compton database detailed in
Ref.~\cite{Griesshammer:2012we}. The results, given in
eq.~\eqref{eq:scalar-pol-values}, are constrained by the Baldin sum rule
$\alphaep+\betamp=13.8\pm0.4$~\cite{Olmos:2001}, though fitting $\alphaep$ and
$\betamp$ separately produces results which are consistent within the
uncertainties. In the same work, it was found that a good fit to (unpolarised)
proton Compton scattering data at $\calO(e^2\delta^4)$ in the amplitudes could
only be achieved if one also determines $\gammamm^{\text{(p)LEC}}$. This makes
$\gammammp$ also a fitted quantity.

The neutron scalar polarisabilities are obtained from deuteron targets.  In
Ref.~\cite{Myers:2014ace}, we extracted scalar polarisabilities for the
neutron from the statistically-consistent world data, updated with the recent
high-quality data from MAX-lab~\cite{Myers:2015aba}, with the neutron's Baldin
sum rule $\alphaen+\betamn=15.2\pm0.4$ as a constraint~\cite{Levchuk:1999zy}.
This extraction was carried out at one order lower, $\calO(e^2\delta^3)$,
which means that the theoretical uncertainties are larger, but there was no
need to fit $\gammammn$.  For the sake of the present study, we take a
minimalist approach and assume that the $\gammamm$ LEC we promoted by one
order is purely isoscalar, while all other short-distance contributions to the
spin polarisabilities enter at higher order.  In Sect.~\ref{sec:results}, we
will show that the dependence of $\gammamm$ on the pion mass provides
supporting evidence for this.

\subsubsection{\texorpdfstring{$\Delta(1232)$}{Delta(1232)} Pole Contribution}

Since $\DeltaM$ is about $30\%$ of $\MN$, recoil effects in this part of the
amplitude are expected to be sizeable. We thus choose to include purely
kinematic, relativistic effects in the Delta pole contribution (see leftmost
graph of Fig.~\ref{fig:Deltapole}). The results in Lorentz-covariant
kinematics were implicit in Ref.~\cite{McGovern:2012ew}, but are stated here
for the first time. Related results, but with several errors, were given in
Ref.~\cite{Pascalutsa:2003zk}.
\begin{align}
  \label{eq:abDelta}
  \begin{split}
    \alphae^{\Delta} &= -\frac{2\alphaEM b_2^2}{9\MN^2(2\MN+\DeltaM)} \\
    \betam^{\Delta} &= \frac{2\alphaEM b_1^2}{9\MN^2\DeltaM}
  \end{split}\\
  \begin{split}
    \label{eq:gDelta}
    \gammaee^{\Delta}&= \frac{\alphaEM}{18\MN^3}\left[
      -\frac{b_1^2}{\DeltaM}+\frac{b_1b_2}{2\MN+\DeltaM}-
      \frac{2b_2^2(\MN+\DeltaM)}{(2\MN+\DeltaM)^2}\right]\\
    \gammamm^{\Delta}&= \frac{\alphaEM}{18\MN^3}\left[
      \frac{2b_1^2(\MN+\DeltaM)}{\DeltaM^2}-\frac{b_1b_2}{\DeltaM}+
      \frac{b_2^2}{2\MN+\DeltaM}\right]\\
    \gammaem^{\Delta}&= \frac{\alphaEM}{18\MN^3}\left[
      -\frac{b_1^2}{\DeltaM}+\frac{3b_1b_2}{2\MN+\DeltaM}\right]\\
    \gammame^{\Delta}&= \frac{\alphaEM}{18\MN^3}\left[
      -\frac{3b_1b_2}{\DeltaM}+\frac{b_2^2}{2\MN+\DeltaM}\right]
  \end{split}
\end{align}
These contributions reduce at leading order in the heavy-baryon limit,
$\MN\to\infty$ and $b_2\to0$, to the results by Hemmert et al.~\cite{hhk97,
  hhkk98}, namely $\gammamm^{\Delta}=\betam^{\Delta}/(2\Delta_M)$. All other
polarisabilities are zero in this limit.  Here, $b_1$ and $b_2$ are the
magnetic and electric $\gamma$N$\Delta$ couplings respectively. Their leading
chiral loop corrections were derived in Ref.~\cite{McGovern:2012ew} but enter
an order of $\mpi/\lambdachi \sim \delta^2$ higher, and are thus omitted from
the results above.

\subsubsection{\texorpdfstring{$\pi \Delta$}{pi-Delta} Loops}

\begin{figure}[!htb]
  \begin{center}
    \includegraphics[width=0.9\textwidth]{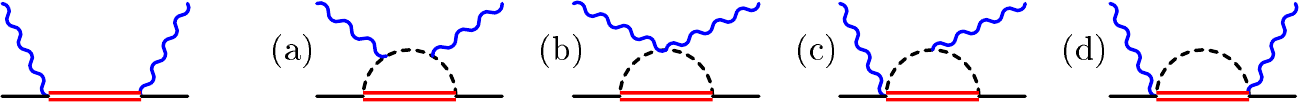}
    \caption{(Colour online) Leading contributions to the polarisabilities
      from the $\Delta(1232)$ (leftmost) and its pion cloud.  Notation as in
      Fig.~\ref{fig:LOpiN}; covariant $\gamma$N$\Delta$ and heavy-baryon
      $\pi$N$\Delta$ vertices from Ref.~\cite{McGovern:2012ew}. Permuted and
      crossed diagrams not displayed.}
    \label{fig:Deltapole}
    \label{fig:piDelta}
  \end{center}
\end{figure}

The leading contributions from the pion cloud around the Delta,
Fig.~\ref{fig:piDelta}, were calculated in Refs.~\cite{hhk97, hhkk98} using
the heavy-baryon approximation:
\newcommand{\arcsinhx}{\;F\!\left(\frac{\DeltaM}{\mpi}\right)}
\begin{align}
  \label{eq:abpiDelta}
  \begin{split}
    \alphae^{\pi\Delta} &= \frac{\alphaEM\gpiNDelta^2}{54\pi^2\fpi^2}
    \left[\frac{9\DeltaM}{\DeltaM^2-\mpi^2}+
      \frac{\DeltaM^2-10\mpi^2}{(\DeltaM^2-\mpi^2)^\frac{3}{2}}\arcsinhx\right]\\
    \betam^{\pi\Delta} &= \frac{\alphaEM\gpiNDelta^2}{54\pi^2\fpi^2}
    \frac{1}{(\DeltaM^2-\mpi^2)^\frac{1}{2}}\arcsinhx\\
  \end{split}&\\
  \label{eq:gpiDelta}
  \begin{split}
    \gammaee^{\pi\Delta} &= \frac{\alphaEM\gpiNDelta^2}{108\pi^2\fpi^2}
    \left[\frac{\DeltaM^2+5\mpi^2}{(\DeltaM^2-\mpi^2)^2}+
      \frac{\DeltaM(\DeltaM^2-7\mpi^2)}{(\DeltaM^2-\mpi^2)^\frac{5}{2}}
      \arcsinhx\right]\\
    \gammamm^{\pi\Delta} &= -\gammaem^{\pi\Delta}= -\gammame^{\pi\Delta} = -
    \frac{\alphaEM\gpiNDelta^2}{108\pi^2\fpi^2}\!
    \left[\frac{1}{\DeltaM^2-\mpi^2}-
      \frac{\DeltaM}{(\DeltaM^2-\mpi^2)^\frac{3}{2}}\!\arcsinhx\right]\!,
  \end{split}&
\end{align}
with $F(x)=\mathrm{arcsinh}(\sqrt{x^2-1})$ and $\gpiNDelta$ the $\pi$N$\Delta$
coupling constant.  These results are real, continuous and non-singular for
all ratios $\mpi/\DeltaM>0$.

\section{Polarisabilities and Uncertainties at the Physical Point}
\setcounter{equation}{0}
\label{sec:physpoint}
\subsection{Summary}\label{sec:values}

\begin{figure}[!b]
  \begin{center}
    \includegraphics[height=0.4\linewidth,clip=]
    {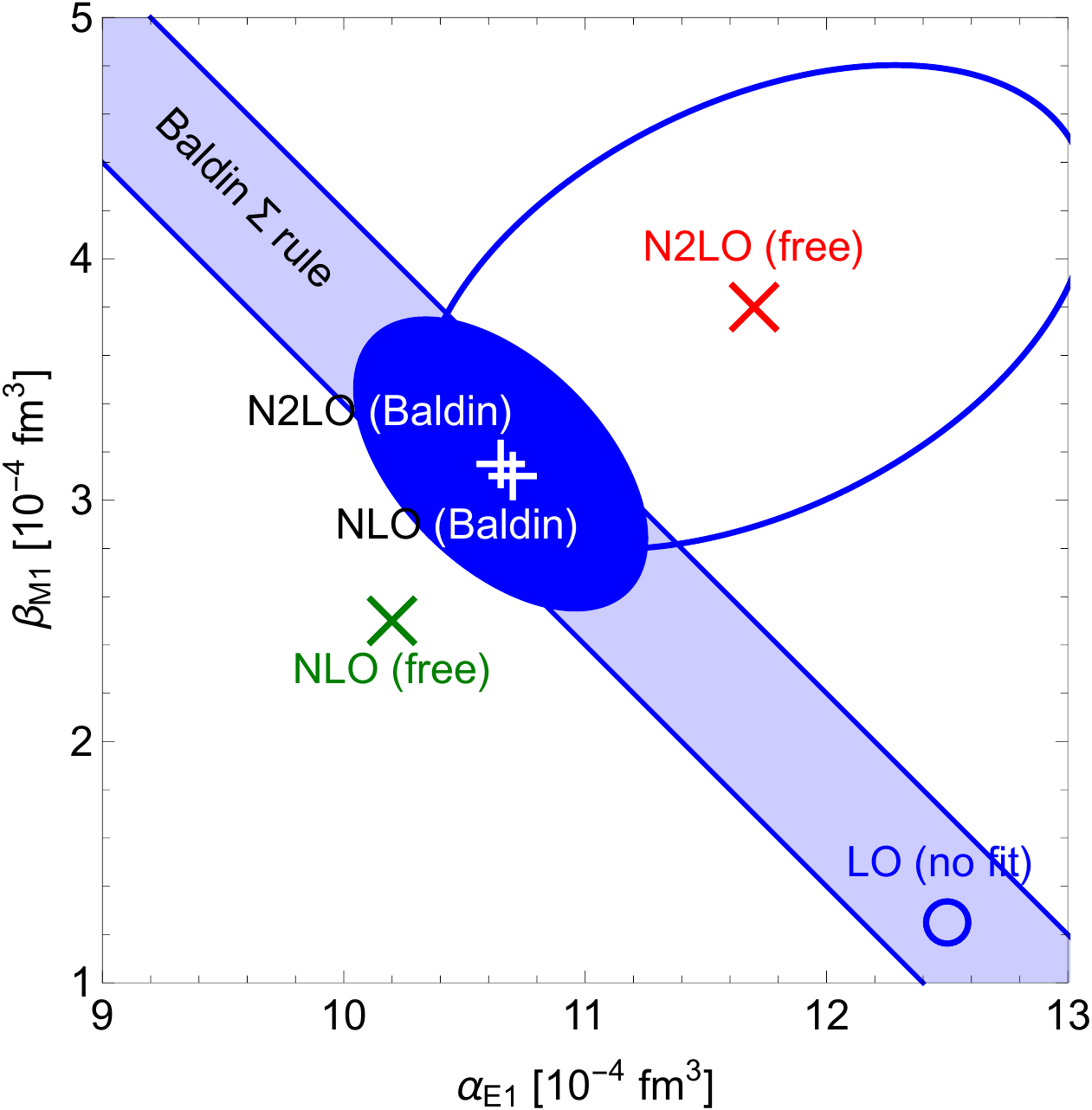}
    \hqqq \includegraphics[height=0.4\textwidth, clip=]{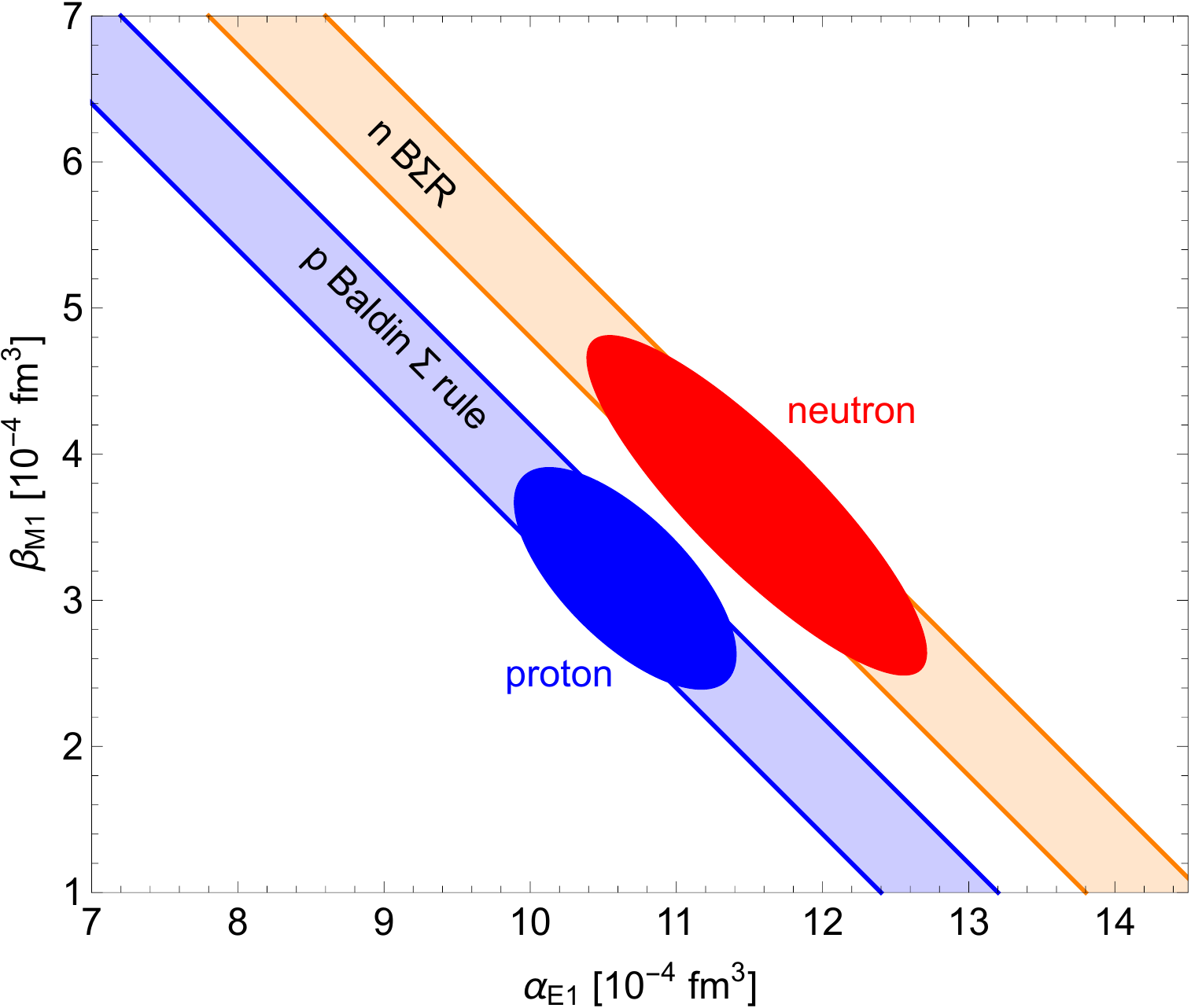}
    \caption{(Colour online) Left: Convergence pattern of the scalar
      polarisabilities of the proton discussed in Ref.~\cite{McGovern:2012ew},
      with and without the Baldin sum rule constraint. Ellipses and lines of
      the \NXLO{2} determination denote $1\text{-}\sigma$ confidence intervals
      and account only for experimental errors. Right: Proton and neutron
      scalar polarisabilities in recent \ChiEFT
      extractions~\cite{McGovern:2012ew, Myers:2014ace}. Ellipses denote
      $1\text{-}\sigma$ confidence intervals and add theoretical,
      experimental/statistical and Baldin-sum-rule related uncertainties in
      quadrature.}
    \label{fig:abvalues}
  \end{center}
\end{figure}

The formulae of the previous section with the LECs for $\alphaep$, $\betamp$,
$\alphaen$, $\betamn$ and $\gammamm^\text{(s)}$ fitted to unpolarised Compton
scattering data give the values of the scalar polarisabilities of the nucleons
in units of $10^{-4}~{\rm fm}^3$ as follows~\cite{McGovern:2012ew,
  Myers:2014ace}:
\begin{align}\label{eq:scalar-pol-values}
  \begin{split}
    \alphaep=&10.65\pm0.35(\text{stat})\pm0.2(\text{Baldin})
    \pm0.3(\text{theory})\\
    \betamp =&\phantom{0}3.15\mp0.35(\text{stat})\pm0.2(\text{Baldin})
    \mp0.3(\text{theory})\\ \alphaen=&11.55\pm1.25(\text{stat})\pm0.2(\text{Baldin})\pm0.8(\text{theory})\\
    \betamn=&\phantom{0}3.65\mp1.25(\text{stat})\pm0.2(\text{Baldin})
    \mp0.8(\text{theory})\;\;,
  \end{split}
\end{align}
with $\chi^2=113.2$ for $135$ degrees of freedom for the proton, and $45.2$
for $44$ for the neutron. Notice that due to the imposition of the Baldin sum
rule for each nucleon, both the statistical and theory errors are
anticorrelated between $\alphae$ and $\betam$; see also
Sect.~\ref{sec:Finding-Theory-Uncertainties}.

For the spin polarisabilities, in units of $10^{-4}~{\rm fm}^4$:
\begin{align}\label{eq:spin-pol-values}
  \gammaeep&=-1.1\pm1.9(\text{theory})\hqq\hqq&\hqq\hqq
  \gammaeen&=-4.0\pm1.9(\text{theory})\nonumber\\
  \gammammp&= \phm2.2\pm0.5(\text{stat})\pm0.6(\text{theory})&
  \gammammn&=\phm1.3\pm0.5(\text{stat})\pm0.6(\text{theory})\nonumber\\
  \gammaemp&=-0.4\pm0.6(\text{theory})&
  \gammaemn&=-0.1\pm0.6(\text{theory})\nonumber\\
  \gammamep&=\phm 1.9\pm0.5(\text{theory})&
  \gammamen&=\phm2.4\pm0.5(\text{theory})
\end{align}
The central values for the proton were given in Ref.~\cite{McGovern:2012ew}
and cited, with an estimation of uncertainties supplied by the current
authors, in Refs.~\cite{talkMAMI, Martel:2014pba}\footnote{The errors of
  $\{\pm1.8;\pm0.7;\pm0.4;\pm0.4\}$ cited in Refs.~\cite{talkMAMI,
    Martel:2014pba}, though supplied by us, differ slightly from these.  That
  is because eq.~\eqref{eq:spin-pol-values} reflects the difference in
  power-counting for amplitudes and polarisabilities discussed in
  Sect.~\ref{sec:pols}; considers isoscalar and isovector convergence
  separately; and uses the Bayesian framework described below to calculate
  68\% intervals from the EFT truncation error. With the possible exception of
  $\gammaem$, we do not consider the change in uncertainties
  significant---cf.~the discussion below of how well uncertainties can be
  estimated.}.

A justification of the theoretical uncertainties in
eqs.~\eqref{eq:scalar-pol-values} and~\eqref{eq:spin-pol-values}, which are
derived from order-by-order convergence of the results, will be the subject of
the next two subsections. Here, we only remark that they encompass $68$\%
intervals but that the corresponding probability is not distributed
in a Gau\3ian manner. Note also that the statistical error from fitting
$\gammammp$ along with $\alphaep$ and $\betamp$ in Ref.~\cite{McGovern:2012ew}
is inherited by related quantities, including $\gammammn$;
cf.~Sect.~\ref{sec:CTs}.

Figure~\ref{fig:abvalues} illustrates the pattern of convergence and the
$1\sigma$ ellipses for the scalar polarisabilities. At present, there is only
a weak signal that proton and neutron polarisabilities differ, and then only
if the Baldin sum rule is used. The ellipses are obtained by adding all
uncertainties (statistical, Baldin, and truncation) in quadrature. The
truncation error plays a minor role, so although, strictly speaking, its
non-Gau\3ian distribution (see Sect.~\ref{sec:theoryerrors} below) mandates a
more sophisticated treatment of error combination, that would not lead to
ellipses which are appreciably different from those shown here.

\begin{table}[!ht]
  \centering
  \begin{tabular}{|l|l|llll|}
    \hline\rule[-1.5ex]{0ex}{4.5ex}
    & this work:
    $\calO(e^2\delta^{-2})$&NLO
    B$\chi$PT & MAMI & DR(I) & DR(II)\\\hline\rule[-1ex]{0ex}{4ex}
    $\gammaeep$&$-1.1\pm1.9$& $-3.3\pm 0.8$&$-3.5\pm1.2$   &
    $-3.4$&$-4.3$ 
    \\\rule[-1ex]{0ex}{4ex}
    $\gammammp$&
    $\phm2.2\pm0.5(\text{stat})\pm0.6$&$\phm2.9\pm 1.5$&$\phm3.2\pm0.9$&
    $\phm2.7$&$\phm2.9$
    \\\rule[-1ex]{0ex}{4ex}
    $\gammaemp$&$-0.4\pm0.6$&$\phm0.2\pm 0.2 $&$-0.7\pm1.2$&
    $\phm0.3   
    $&$\phm0.0$ 
    \\\rule[-1ex]{0ex}{4ex}
    $\gammamep$&$\phm1.9\pm0.5$&$\phm1.1\pm0.3$&$\phm2.0\pm0.3$&
    $\phm1.9$&$\phm2.2$
    \\\hline\rule[-1ex]{0ex}{4ex} $\gammazerop$&$-2.6\pm0.5(\text{stat})\pm1.8$&$-0.9\pm1.4$&$-1.0\pm0.1\pm0.1$&
    $-1.5$&$-0.8$ 
    \\\rule[-1ex]{0ex}{4ex}
    $\gammapip$&$\phm5.5\pm0.5(\text{stat})\pm1.8$&$\phm7.2\pm1.7$&$\phm8.0\pm1.8$& 
    $\phm7.8$&$\phm9.4$
    \\\hline
  \end{tabular}
  \caption{Values of the proton spin polarisabilities from the current
    calculation, from covariant $\chi$PT at
    NLO~\cite{Lensky:2015awa}, from experiment~(``MAMI'' \cite{Martel:2014pba};
    incorporating $\gammazero$:~\cite{Ahrens:2001qt, Dutz} and $\gammapi$:~\cite{Camen}); and from Dispersion
    Relations (I)~\cite{Babusci:1998ww}, (II)~\cite{Holstein:1999uu,
      Hildebrandt:2003fm, Pasquiniprivcomm, Martel:2014pba}.
\label{tab:spinpols}  
}
\end{table}

Table \ref{tab:spinpols} gives the comparison with the recent MAMI extraction
of the proton spin polarisabilities from the first double-polarised Compton
measurements~\cite{Martel:2014pba}. These extractions were constrained to
reproduce the listed values of two linear combinations, namely the forward and
backward spin polarisabilities $\gamma_0$~\cite{Ahrens:2001qt, Dutz,
  Pasquini:2010zr} and $\gamma_\pi$~\cite{Camen}.  (Note, however, that the
value for $\gammazero$ is much more model-independent than that for
$\gammapi$.) Within their stated uncertainties, they all overlap our $68$\%
confidence intervals.  This agreement, and specifically the agreement for
$\gammammp$, supports our previously-developed strategy of promoting and
fitting the isoscalar LEC of $\gammamm$ to unpolarised
data~\cite{McGovern:2012ew}.  Future experiments on double-polarisation
observables on the proton and light nuclei that are running or approved at
MAMI and \HIGS will provide high-accuracy data for more conclusive
comparisons~\cite{Weller:2009zz, HIGSPAC, Downie:2011mm, Huber:2015uza}. A
\ChiEFT analysis of the polarised scattering data for the proton is
forthcoming~\cite{observables}.

At this order, the only differences between neutron and proton spin
polarisabilities come from the pion-cloud corrections of
eq.~\eqref{eq:gpiN-corr}.  Little is known about the neutron spin
polarisabilities. Our prediction $\gammazeron=0.5\pm0.5(\mathrm{stat}) \pm1.8$
is certainly compatible with the expectation that this quantity should be
``about zero''~\cite{Schumacher:2005an}. For $\gammapin$, our value is
$7.7\pm0.5(\mathrm{stat})\pm1.8$. This places us on the low end of the range
$12.7\pm4.0$ extracted in a DR framework from the reaction
$\gamma\text{d}\to\gamma {\rm p} {\rm n}$ \cite{Kossert:2002ws}, but
uncertainties in that theory may be underestimated~\cite{Griesshammer:2012we}.

\subsection{A Theory Of Theoretical Uncertainties}
\label{sec:theoryerrors}

The intrinsic uncertainty of any EFT calculation comes from the truncation of
the EFT series at a finite order $k$. It is clear that definitive results for
terms in the series that have not been computed are impossible to obtain.
Nevertheless, estimates of the EFT truncation error can be made using the same
strategy as in other perturbative quantum field theories, i.e., by combining
knowledge of the perturbative parameter with reasonable assumptions about the
behavior of higher-order coefficients.  Ref.~\cite{Cacciari:2011ze} used
Bayesian methods to implement this strategy for perturbative QCD, and
Ref.~\cite{Furnstahl:2015rha} adapted the approach to EFTs. Other methods for
estimating the truncation error are certainly possible, but this Bayesian
framework has the advantage that it allows clear specification of premises: it
facilitates a rigorous derivation of the theoretical uncertainties from a
particular assumption about the behaviour of higher-order coefficients in the
EFT series.  Note that we work in the leading-omitted-term approximation,
i.e.~we assume the error associated with that term dominates the theoretical
uncertainty. As long as \ChiEFT is convergent, this is a reasonable assumption
(cf.~discussion of its accuracy below). We also point out that Bayesian
methods can aid in the extraction and uncertainty estimation of \ChiEFT LECs
that appear in the nucleon polarisabilities~\cite{Schindler:2008fh,
  Wesolowski:2015fqa}, but we do not pursue that avenue here, instead applying
them only to truncation errors.

\def\scale{\ensuremath{R}} 
\def\error{\ensuremath{\Delta}} 
\def\centralvaluexi{\ensuremath{\xi_{(k)}}}
\def\scalexi{\ensuremath{\scale_\xi}} \def\errorxi{\ensuremath{\error}}

Suppose we have computed $k$ non-trivial orders of a generic polarisability
$\xi$, with the first (leading) order being $c_0$:
\begin{equation}
  \centralvaluexi= \sum_{n=0}^{k-1} c_n \delta^n\;\;.
  \label{eq:xiexp}
\end{equation}
The canonical EFT estimate of the truncation error, \scalexi, is then
determined by the largest magnitude of the coefficients $c_n$ as follows:
\begin{equation}
  \scalexi=\max_n \{|c_n|: n=0, \ldots, k-1\}\times\delta^k \;\;.
  \label{eq:errorestimate}
\end{equation}
Equation~\eqref{eq:errorestimate} was the basis of the determination of the
truncation error of scalar polarisabilities in
Refs.~\cite{Griesshammer:2012we,McGovern:2012ew}. An entirely equivalent error
estimation was recently articulated and advocated for two- and
three-nucleon-system observables in Refs.~\cite{Epelbaum:2014,
  Epelbaum:2014sza,Binder:2015mbz}.

We want to understand how to interpret such a truncation error. To do that, we
compute the probability distribution function (pdf), denoted
$\pr(\errorxi|I)$, namely the degree of belief that a polarisability will take
a specific value which differs by an amount $\error$ from the calculated
central value $\centralvaluexi$ of the \ChiEFT prediction at order $k$. This
belief is based upon the available information $I$ which includes the order
$k$ of the calculation, the behaviour of the \ChiEFT series, our expectations
regarding naturalness, and---in the case of fitted polarisabilities---data at
the physical pion mass.  There is no reason to assume this pdf will be
Gau\3ian; in general, it is not.  Nevertheless, it can still be integrated to
compute degree-of-belief intervals (DoB intervals), such as the Bayesian
analogue of the usual 68\% (1$\sigma$) and 95\% (2$\sigma$) confidence
intervals.

Since this is a Bayesian approach, a choice of prior for the coefficients
$c_j$ associated with the higher-order terms is mandatory.  We implement a
prior which assigns uniform probability to any value of the omitted
coefficients, up to some unspecified maximum, as for prior $A_\epsilon^{(1)}$
of Ref.~\cite{Furnstahl:2015rha}. These assumptions lead to analytic
expressions for the probability distribution:
\begin{equation}
  \text{pr}(\errorxi | \scalexi,k)=\frac{k}{k+1}\;
  \frac{1}{2\scalexi}\times\left\{\begin{array}{ll}
      1&\mbox{ for } |\errorxi | \le \scalexi\\
      \left(\dis\frac{\scalexi}{ |\errorxi|}\right)^{k+1}
      &\mbox{ for } |\errorxi | > \scalexi\end{array}\right.
  \label{eq:theoryerror}
\end{equation}
For $k=1,2,3$, these pdfs are illustrated for in
Fig.~\ref{fig:combinedpdfs}. By integration of the pdf, we define $\sigma_\xi$
such that $[\centralvaluexi-\sigma_\xi;\centralvaluexi+\sigma_\xi]$ is the
68\% DoB interval. This is also how we define the theoretical or truncation
error in eqs.~\eqref{eq:scalar-pol-values} and \eqref{eq:spin-pol-values}, and
throughout this work. For comparison, the standard EFT error estimate,
$[\centralvaluexi-\scalexi;\centralvaluexi+ \scalexi]$, is a $\frac{k}{k+1}
\times 100$\% DoB interval. For two of the cases we are concerned with,
namely~$k=2$ and $k=3$, these are $67\%$ and $75\%$ intervals, respectively,
so there is little difference between $\scalexi$ and $\sigma_\xi$.

We close with a few comments. First, the truncation uncertainty
\eqref{eq:theoryerror} is not distributed in a Gau\3ian way.  This can be
understood as follows. Each time a new order is calculated, more information
is gleaned about the largest-possible coefficient in the series. That the
probability is equidistributed for any value inside the standard EFT interval
between zero and $R_\xi$ is inherited from our choice of prior. On the other
hand, the probability of finding a coefficient larger than the maximum of
those obtained thus far becomes smaller the more orders are known---a fact
represented by the steeper power-law falloff above the maximum as the EFT
calculation's order increases. Indeed, the $95\%$ DoB interval of the pdf in
eq.~\eqref{eq:theoryerror} is not twice as large as the $68$\% interval
$\sigma_\xi$, as would be expected for a Gau\3ian pdf; see also the examples
in Fig.~\ref{fig:combinedpdfs} below. Instead, it lies at about $7\sigma$ for
$k=1$, $2.6\sigma$ for $k=2$, and $1.9\sigma$ for $k=3$. Second, in cases
where we know that one of the $c_n$ is {\it a priori} zero, e.g.~by symmetry
arguments, since the power-law falloff of the pdf outside the ``standard'' EFT
interval is determined by the number of non-trivial orders computed, the value
of $k$ to be used in eq.~\eqref{eq:theoryerror} must then be reduced by one.

Different interpretations of the naturalness of EFT coefficients are encoded
in different priors, and those in turn can produce somewhat different 68\% DoB
intervals. In Ref.~\cite{Furnstahl:2015rha}, several possibilities for the
naturalness prior were considered and it was demonstrated that---once three
orders in the EFT series are known---those different interpretations of
naturalness led to a 10--15\% variation in the truncation error. However,
the variation is larger if fewer coefficients in the series have
been computed, as then the specific form of the prior plays more of a
role in determining the final error.  Furthermore, our invocation of the
first-omitted-term approximation means that our error bands have a fractional
uncertainty that could be as large as ${\mathcal O}(\delta)$, although the
actual impact of terms beyond the (first omitted) $\delta^k$ term of
Eq.~(\ref{eq:errorestimate}) depends on whether coefficients at order $k$ and
beyond are correlated, uncorrelated, or
anti-correlated~\cite{Furnstahl:2015rha}. Regardless, this suggests that
the truncation error computed here should be understood to itself have an
accuracy of $\pm 20$\%. The errors quoted below should be read with this in
mind. For uniformity of presentation, we quote all errors to one decimal
place, but in certain instances this may constitute spurious precision.
 
\subsection{Assigning Theoretical Uncertainties to Spin Polarisabilities}
\label{sec:Finding-Theory-Uncertainties}

The values of the spin polarisabilities at $\calO(e^2\delta^{-4})$ and
$\calO(e^2\delta^{-2})$ are given in Table~\ref{tab:convergence}. The first
order is isoscalar, while the next one contains both isoscalar and isovector
components. The series for the isoscalar spin polarisabilities is therefore
computed to an accuracy of $k=2$ nonzero orders; the contribution of order
$\delta^1$ relative to LO is zero, and so we know only $c_0$ and $c_2$ in
eq.~\eqref{eq:errorestimate}. We hence obtain the results for the isoscalar
remainder $\scalexi$ from eq.~\eqref{eq:errorestimate} shown in
Table~\ref{tab:convergence}.  In practice $c_2$ is the larger coefficient in
each case, and so $\scale_{\gamma_i}$ is $\delta$ times the $e^2\delta^{-2}$
contribution. Comparing with eq.~\eqref{eq:theoryerror} for $k=2$, we find
that this remainder can be interpreted as a $67$\% DoB interval.  For the
isovector spin polarisabilities we only have one order, and so
$\scale_{\gamma_i}$ is $\delta$ times the total and the corresponding DoB
interval is only $50$\%.

\begin{table}[!ht]
  \centering
  \begin{tabular}{|l|llll|}
    \hline
    \rule[-1.5ex]{0ex}{4.5ex}
    &$\gammaee$&$\gammamm$&$\gammaem$&$\gammame$\\\hline
    isoscalar $e^2\delta^{-4}$& $-5.7$&$-1.1$&$\phantom{+}1.1$&$\phantom{+}1.1$\\
    isoscalar
    $e^2\delta^{-2}$&$-2.6\pm1.3$&$\phantom{+}1.8
    \pm0.5$&$-0.3\pm0.6$ &$\phantom{+}2.2\pm0.4$ \\\hline
    isovector $e^2\delta^{-2}$&$\phantom{+}1.5\pm0.6$& $\phantom{+}0.5\pm0.2$& $-0.1\pm0.1$& $-0.2\pm0.1$\\\hline 
  \end{tabular}
  \caption{Predictions of the spin polarisabilities at each \ChiEFT
    order. The total is given, so that the isoscalar \emph{addition} at
    $\calO(e^2\delta^{-2})$  is the difference between the first and second
    lines. At $\calO(e^2\delta^{-2})$, the canonical EFT uncertainty estimates
    $\scale_{\gamma_i}$ of eq.~\eqref{eq:errorestimate} is shown as an error on the central value (in this table only, this ``error'' is not the $68$\% DoB interval).\label{tab:convergence}}
\end{table}

\begin{figure}[!htb]
  \begin{center}
    \includegraphics[width=0.8\textwidth]{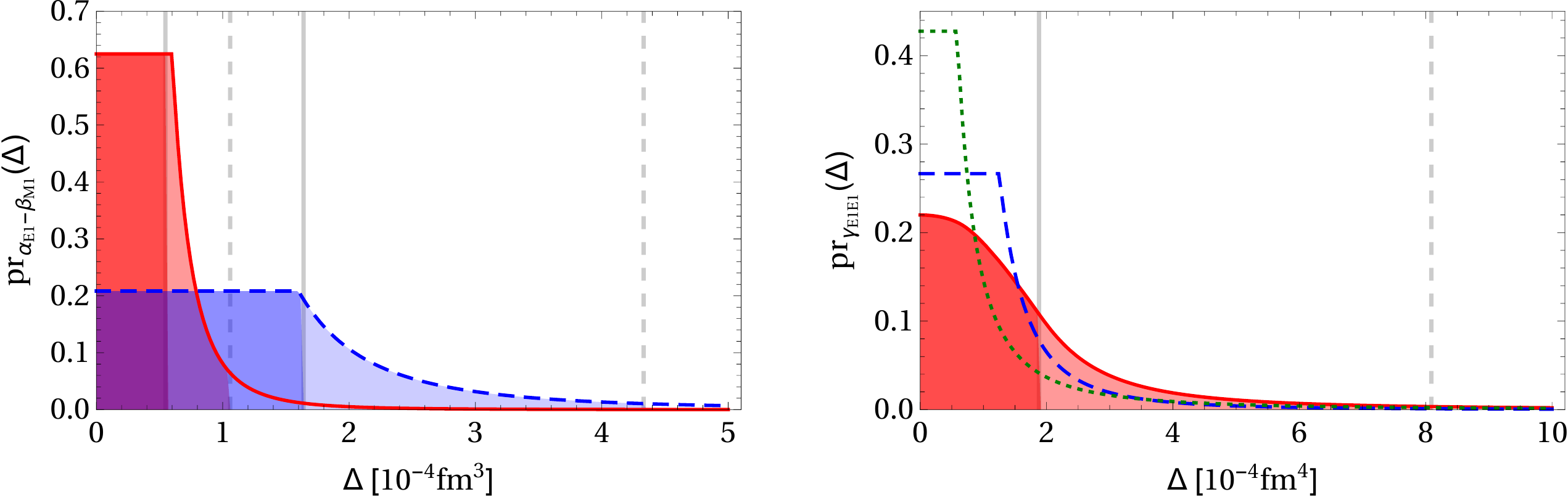}
\caption{(Colour online) Examples of pdfs. Left: The pdf of
  $\alphae-\betam$ at the physical point for the proton (red solid) and
  neutron (blue dashed) as resulting from the fits underlying
  eq.~\eqref{eq:scalar-pol-values}, at \NXLO{2} for the proton ($k=3$,
  $\scale_{\alphae-\betam}=0.6$ in eq.~\eqref{eq:theoryerror}) and NLO for the
  neutron ($k=2$, $\scale_{\alphae-\betam}=1.6$).  Right: The pdf of the
  truncation error in $\gammaee$ at the physical point. The green dotted line
  shows the isovector combination's truncation error ($k^{(\text{v})}=1$, $
  \scale{}^\text{(v)}_{\gammaee}=0.6$), the blue dashed line that of the
  isoscalar ($k^{(\text{s})}=2$, $ \scale{}^\text{(s)}_{\gammaee}=1.3$), and
  the red line is the pdf that results from the integration in
  eq.~\eqref{eq:combining}.
  The solid (dashed) grey line denotes the 68\% (95\%) DoB interval. Pdfs for
  all polarisabilities are available in the Appendix.}
\label{fig:combinedpdfs}
\end{center}
\end{figure}

To get the pdf of the truncation error for an individual nucleon, we convolute
the two pdfs:
\begin{equation}
  \text{pr}_\xi(\errorxi)\equiv\text{pr}(\errorxi | \scale{}^\text{(s)}_\xi,k^\text{(s)},
  \scale{}^\text{(v)}_\xi,k^\text{(v)})
  =\int\limits_{-\infty}^\infty\dd y\,
  \text{pr}(y| \scale{}^\text{(s)}_\xi,k^\text{(s)})
  \text{pr}(\errorxi-y|\ \scale{}^\text{(v)}_\xi,k^\text{(v)})\;\;,
  \label{eq:combining}
\end{equation}
with obvious notation for isovector and isoscalar pieces. As
Fig.~\ref{fig:combinedpdfs} shows, this smears out the individual pdfs
considerably so that they look somewhat more Gau\3ian. Results are the same
for the proton and neutron.  The additional convolution makes the relation
between $95$\% and $68$\% DoBs depend on both $\errorxi^\text{(s)}$ and
$\errorxi^\text{(v)}$.

In all cases, simply adding isoscalar and isovector errors linearly yields
ranges that are near-identical to the 68\% DoB intervals.  One could also
choose to study proton and neutron convergence patterns separately, instead of
those of the isoscalar and isovector quantities. In general this produces
somewhat smaller errors; it never gives larger ones. For example, the proton
uncertainty of $\gammaee$ would be about equally large, while that for the
neutron would be quite a bit smaller ($\pm0.7$). Whether one uses the
proton-neutron or the isospin basis is a question of choice. We take the
latter because its error assessments are more conservative and because \ChiEFT
is most naturally formulated in the isospin basis.

We now turn to $\gammamm$ since its expansion is more complicated. As
mentioned above, it is, strictly speaking, not a free parameter at the order
to which we work, but in practice its proton value was obtained from a fit to
unpolarised proton data.  This requires the promotion of a LEC from the
fifth-order $\pi$N Lagrangian.  A further complication is that its Delta-pole
contribution of $+2.8$, which is nominally suppressed by $\delta^2$, is more
than twice the LO contribution. The contribution of $\pi\Delta$ loops is tiny.
The origin of this, of course, is the large size of the magnetic
$\gamma$N$\Delta$ coupling $b_1$, whose square enters the leading pole
contribution for $\gammamm$ but not the other spin polarisabilities (see
eq.~\eqref{eq:gDelta}).  As a result, the pole contribution is about $6$ times
as large as if $b_1$ were of natural size.  There is precedent for promoting
terms involving coefficients which are unnaturally large, see
e.g.~\cite{Griesshammer:2000mi}, and hence for $\gammamm$ we treat the $b_1^2$
contribution to the Delta pole as suppressed by only one power of $\delta$,
giving $k=3$ nontrivial orders in our expansion. The calculation is still
complete to \NXLO{2} as any other graphs involving $b_1^2$ will be suppressed
by $\delta^4$; thus only the error estimate is affected by the
reordering. With neither of these adjustments, we would predict the proton
value to be $6.4 \pm 3.0$, which is in fact still compatible with our quoted
result of $2.2\pm0.5(\text{stat})\pm0.6(\text{theory})$.  However several
lines of evidence, though not conclusive, suggest the latter is a more
appropriate central value and uncertainty; these include the close similarity
of the values in the third and fourth-order extractions of the scalar
polarisabilities discussed in Ref.~\cite{McGovern:2012ew}, as well as the DR
and MAMI values quoted in Table~\ref{tab:spinpols}.

Finally, the forward and backward spin polarisabilities of
eq.~\eqref{eq:fbspin} are not independent of the multipole spin
polarisabilities. Their errors could thus be assessed naively as $\pm2.1$ by
adding the individual multipole errors in quadrature, if the individual errors
were Gau\3ian distributed. Since both also inherit the large Delta-pole
contribution as well as the fitted LEC from $\gammamm$, we instead assess
their theoretical uncertainties by the same prescription as for $\gammamm$. As
both are linear combinations of quantities which are sensitive to different
multipolarities, and hence different physical mechanisms, {\it a priori} we
expect the two errors to be similar. Surprisingly, we find an error of
$\gammapi$ that is half that of $\gammazero$. We see no physical reason for
this. Furthermore, when in Sect.~\ref{sec:errorregimei}, we look at values of
$\mpi$ greater than the physical value, the difference between the
$\gammazero$ and $\gammapi$ errors disappears for $\mpi\gtrsim170\;\MeV$. This
suggests that the rather small uncertainty for $\gammapi$ found via the
order-by-order prescription at $\mpiphys$ is accidental. Therefore, for all
$\mpi$, we will use as the error of $\gammapi$ that derived by Bayesian
criteria for $\gammazero$.  The resulting $\gammapip$ uncertainty of $\pm1.8$
at $\mpiphys$ is then close to the added-in-quadrature result.

We close with remarks which pertain to scalar polarisabilities; further
discussion can be found in Sect.~\ref{sec:errorregimei}. Our
previously-published uncertainties of eq.~\eqref{eq:scalar-pol-values} were
also obtained by considering order-by-order convergence, though without the
rigorous framework provided here. For the proton, there are three
non-vanishing orders ($k=3$) and the uncertainties should be interpreted as
the $75\%$ DoB interval for the pdf, rather than $68$\% DoB intervals.  For
the neutron, though, $k=2$ for the extracted scalar polarisabilities, so the
DoB is $67$\%, but statistical uncertainties there far outweigh the
theoretical accuracy.  For either nucleon, our previous remainder estimates
$R_\xi$ and the new $68$\%-DoBs $\sigma_\xi$ are identical after rounding. A
determination for the neutron at $\calO(e^2\delta^4)$ in the amplitudes is
forthcoming~\cite{dcompton-delta4}. Figure~\ref{fig:combinedpdfs} shows the
plots of the pdfs for $\alphae-\betam$, the one free variable after the Baldin
sum rule is used as a constraint.

\subsection{Convergence Tests: Selected Higher-Order Terms}
\label{sec:someho}

To check our error estimate at $\mpiphys$, we look at the effect of including
selected, renormali\-sation-group-invariant, higher-order corrections.  This
does not lead to increased accuracy; in general that is only achieved when all
contributions at a given order are known. It can, however, add credence to our
procedure if contributions fall within the derived error estimates. Since
Refs.~\cite{Griesshammer:2012we, McGovern:2012ew, Myers:2014ace} have already
discussed various such effects for the scalar polarisabilities, we here
consider only the spin polarisabilities.

As mentioned in Sect.~\ref{sec:pols}, some contributions are known in variants
which differ by contributions at $\calO(\delta^{3})$ relative to LO, i.e.~at
the first order not calculated. We will also use these variants to discuss the
reliability of our uncertainty estimate of the $\mpi$-dependence in
Sect.~\ref{sec:tests}.
\begin{enumerate}
\item Vertex corrections to $b_1$ and $b_2$ are reported in
  Ref.~\cite{McGovern:2012ew}. They add $0.1$ units to $\gammaee$ and
  $\gammaem$, $-0.37$ to $\gammamm$, and $-0.2$ to $\gammame$, well within our
  proposed theoretical uncertainties.

\item The $\pi\Delta$ contributions in the heavy-baryon version of
  eqs.~\eqref{eq:abpiDelta} and \eqref{eq:gpiDelta} differ by terms of
  $\calO(e^2\delta^{-1})$ from a covariant version~\cite{Lensky:2015awa}. The
  latter add $-0.2$ to $\gammaee$, $-0.4$ to $\gammamm$, $0.3$ to $\gammaem$
  and $0.03$ to $\gammame$, again in accord with our error bars.

\item The difference between the covariant Delta-pole result of
  eqs.~\eqref{eq:abDelta} and \eqref{eq:gDelta} and the heavy-baryon one
  enters at $\calO(e^2\delta^{-1})$ as well. The latter sets the $E2$-coupling
  $b_2$ and recoil effects to zero and uses a different value of
  $b_1^\text{HB}\approx4.8$~\cite{Griesshammer:2012we, McGovern:2012ew}.  We
  find the heavy-baryon version adds $0.35$ units to $\gammaee$, $0.8$ to
  $\gammamm$, $0.4$ to $\gammaem$ and $-0.3$ to $\gammame$. These are within
  the quoted uncertainties.
\end{enumerate}
In each case, the changes quoted for $\gammamm$ do not take into account the
fact that it needs to be refitted once such effects are added; we expect this
to decrease the magnitude of higher-order effects.

\section{Chiral Extrapolations}
\setcounter{equation}{0}
\label{sec:extrapolations}

A chiral extrapolation of the central values for the polarisabilities simply
uses the pion-mass dependence of the \ChiEFT results in Sect.~\ref{sec:pols}
to generate the functions $\xi(\mpi)$. The final results are shown as solid
red (proton) and blue (neutron) lines in all the plots of
Sect.~\ref{sec:results}.

Just as at the physical point, the chiral predictions have truncation errors
which are assessed as before, but now are functions of $\mpi$; we will make a
few technical remarks on their evaluation in Sect.~\ref{sec:errorregimei}. The
truncation uncertainties are represented in the plots by shaded bands for as
far as we trust them; see Sect.~\ref{sec:errorregimeii}. These error bands at
unphysical pion masses correspond only to the ``theory'' error. The full error
is then obtained by combining this with the ($\mpi$-independent) statistical
and (for scalar polarisabilities) Baldin-sum-rule errors.  Therefore, in the
plots, we mark the total error only at the physical point, adding all errors
linearly.  In cases where the physical-point error is larger than the width of
the band, the whole band can be moved up or down within this difference; see
details in Sect.~\ref{sec:results}. This presentation has been chosen to
highlight the running of the error with the pion mass, and because the other
sources of error depend on the current state of the data, and thus may change
in future.  The different sources of errors could of course be combined in a
Bayesian formalism, but that is outside the scope of this paper.

Finally, we do not consider any implicit $\mpi$-dependence of $g_A$, $\fpi$,
$\MN$, $\DeltaM$, $c_i$, etc. All such effects are of higher order in the
chiral counting. For example, the correction to
$g_A(\mpi)=g_A^0+\calO(\mpi^2)$~\cite{Procura:2006gq} is suppressed by two
orders in $\mpi/\Lambda_\chi$ and hence beyond the accuracy of our results.

\subsection{Theoretical Uncertainties near the Physical Point}
\label{sec:errorregimei}

We first consider the uncertainties in regime (i), $\mpi\approx\mpiphys$.
There, we have included those contributions to the polarisabilities that are
required in the physical low-energy Compton amplitudes up to $\calO
(e^2\delta^4)$.  Omitted terms are therefore $\calO (\delta^3)$ relative to
leading. Ignoring a few details to be discussed shortly, we estimate the
remainder $R_\xi(\mpi)$ and $68$\% DoBs $\sigma_\xi(\mpi)$ at any given $\mpi$
in regime (i) just as we did at $\mpiphys$ in
Sect.~\ref{sec:Finding-Theory-Uncertainties}.

The only difference is in the expansion parameter $\delta$, which was defined
in Sect.~\ref{sec:formalism} and enters in eqs.~\eqref{eq:xiexp} and
\eqref{eq:errorestimate}. So long as we remain well below
$\mpi\approx\DeltaM$, it could be taken to be constant, $\delta\approx0.4$.
But, in practice, some chiral corrections to pion loops grow linearly with
$\mpi$, and once $\mpi\approx\DeltaM$ the whole counting changes: Delta graphs
become as important as nucleon ones; see Sect.~\ref{sec:regimes} and the next
subsection.  Hence, we choose a value of $\delta$ which increases with $\mpi$,
\begin{equation}
  \delta(\mpi)=0.4 \frac{\mpi}{\mpiphys}\;\;.
\end{equation}
For the three spin polarisabilities which are pure predictions, $\gammaee$,
$\gammaem$ and $\gammame$, there is little more be said.  As at the physical
point, at any given $\mpi$ we derive the isoscalar and isovector uncertainties
from the order-by-order convergence of their respective series, and obtain the
total uncertainty by convoluting the corresponding pdfs.  The overall scale of
their isoscalar uncertainty $R_{\gamma_i}^{(\text{s})}$ turns out to be
dominated by $\delta$ times the $\calO(e^2 \delta^{-2})$ contribution at all
pion masses we consider.  In each case, the absolute error increases rather
modestly as a function of $\mpi$, mainly because an important part of the
$\calO(e^2 \delta^{-2})$ contribution is actually falling as $1/\mpi$ while
the expansion parameter $\delta(\mpi)$ grows linearly.  The ordering of
contributions to $\gammamm$, $\gammazero$ and $\gammapi$ for
$\mpi\neq\mpiphys$ also mirrors exactly that at the physical point as
discussed in Sect.~\ref{sec:Finding-Theory-Uncertainties}: terms quadratic in
the rather large coupling $b_1^2$ are promoted by one order.

The scalar polarisabilities require a little more discussion, since we did not
derive in detail their errors at the physical point, these having been
determined through fits to data~\cite{McGovern:2012ew,Myers:2014ace}. First,
we recall that the fit of $\alphaen$ and $\betamn$ to deuteron data used
amplitudes only at NLO, while that for the proton used those at one order
higher. This leads to substantially larger truncation errors at the physical
point than that for the proton: $\pm 0.8$ as opposed to $\pm 0.3$. However,
their running with $\mpi$ is predicted in \ChiEFT to \NXLO{2} for both. Thus,
when generating the scalar-polarisability error bands in the neutron case, we
use the same uncertainties as for the proton, while including the larger
truncation uncertainty in the error bar at the physical point.  Once again,
any change in the central value of $\alphaen$ and $\betamn$ resulting from a
new fit to deuteron data at order $\delta^2$ rather than $\delta^1$ relative
to LO only serves to move the curve up and down.  A construction of isoscalar
and isovector uncertainties as for the spin-polarisabilities, with reasonable
assumptions as to how much $\alphaen-\betamn$ may shift in an future \NXLO{2}
fit, produces near-identical bands.

Furthermore, in the proton fits of Ref.~\cite{McGovern:2012ew}, the actual fit
parameter was $\alphaep-\betamp$, with $\alphaep+\betamp$ fixed by the Baldin
sum rule. The scale $\scale$ of the uncertainty was obtained by considering
$\alphaep-\betamp$ as \emph{predicted} at LO---$\calO(e^2\delta^{-2})$---and
\emph{fit} at the next two orders. By fitting the amplitudes at NLO, the LECs
$\alphae^{\text{(p)LEC}}$ and $\betam^{\text{(p)LEC}}$ are promoted by one
order. In practice, the proton fits at orders $\delta^1$ and $\delta^2$
relative to LO gave almost identical values for $\alphae^\text{(p)LEC} -
\betam^\text{(p)LEC}$.  To reflect this, we treat the LECs as part of the
$\calO(e^2\delta^{-1})$ contribution to $\alphaep$ and $\betamp$. The
additional contribution at $\calO(e^2\delta^{0})$ is then subtracted to give
zero at the physical point, and so affects $\alphaep-\betamp$ only for $\mpi
\neq \mpiphys$.

Beyond the physical point, we continue to treat $\alphae+\betam$ and
$\alphae-\betam$ as independent.  However, the Baldin sum rule for
$\alphae+\betam$ is special: at the physical point it provides extrinsic
information in our NLO and \NXLO{2} fits and should be reproduced exactly.  In
fact, higher-order effects in the scalar-polarisability sum for the proton at
the physical point are tiny compared to the LO prediction of $13.9$ from
eq.~\eqref{eq:abpiN-LO}. We therefore base our error estimates for
$\alphaep+\betamp$ beyond the physical point only on the shifts from the LO
term, i.e.~we drop $c_0$ in the construction of $\scalexi$ in
eq.~\eqref{eq:errorestimate}; we then have $k=2$ for $\alphaep+\betamp$.  We
considered several variants of this procedure, but they do not substantially
alter the outcome: while the \ChiEFT uncertainty at $\mpiphys$ is indeed zero,
it grows very rapidly with $\mpi$, see Fig.~\ref{fig:constraints}. A
several-hundred-per-cent error at $\mpi=350\;\MeV$, while technically correct
for a theory defined as an expansion around the chiral limit, seems overly
pessimistic.  Indeed, it defies the quite conservative expectation that
$\alphae+\betam$ tends asymptotically to a constant as $\mpi$ gets bigger,
since the nucleon size, and hence the typical size over which its charged
constituents can fluctuate, will not increase with $\mpi$.  The \ChiEFT error
evolution for $\alphaep - \betamp$ is more modest, and is more representative
of the truncation error in \ChiEFT predictions for polarisabilities.  Well
away from the physical point and the constraint provided by the Baldin sum
rule, there is no particular logic to using $\alphae \pm \betam$, rather than
$\alphae$ and $\betam$ separately, as the quantities via which we assess
convergence of the \ChiEFT series.  Once $\mpi$ reaches $250\;\MeV$, $\alphae$
and $\betam$ separately have much better convergence properties than does
$\alphae + \betam$. For pion masses in that vicinity, taking the error only
from $\alphae-\betam$ gives errors consistent with those obtained from
analysing $\alphae$ and $\betam$ separately. So, in what follows, we simply,
if somewhat arbitrarily, assign zero truncation uncertainty for the evolution
of $\alphae+\betam$ in order to generate our final results for $\alphae$ and
$\betam$.

We then use different colours for the bands in the plot of $\alphae + \betam$
in Fig.~\ref{fig:constraints} to indicate that they are not used to derive
corridors for $\alphae$ or $\betam$.  The widths of the corridors of the
scalar polarisabilities are set by half the uncertainty in $\alphae-\betam$,
and their corridors are anti-correlated. The Baldin-related error is
indicated, along with the statistical one, only at the physical point.

\label{sec:tests}

\begin{figure}[!htb]
  \begin{center}
    \includegraphics[width=0.8\textwidth]{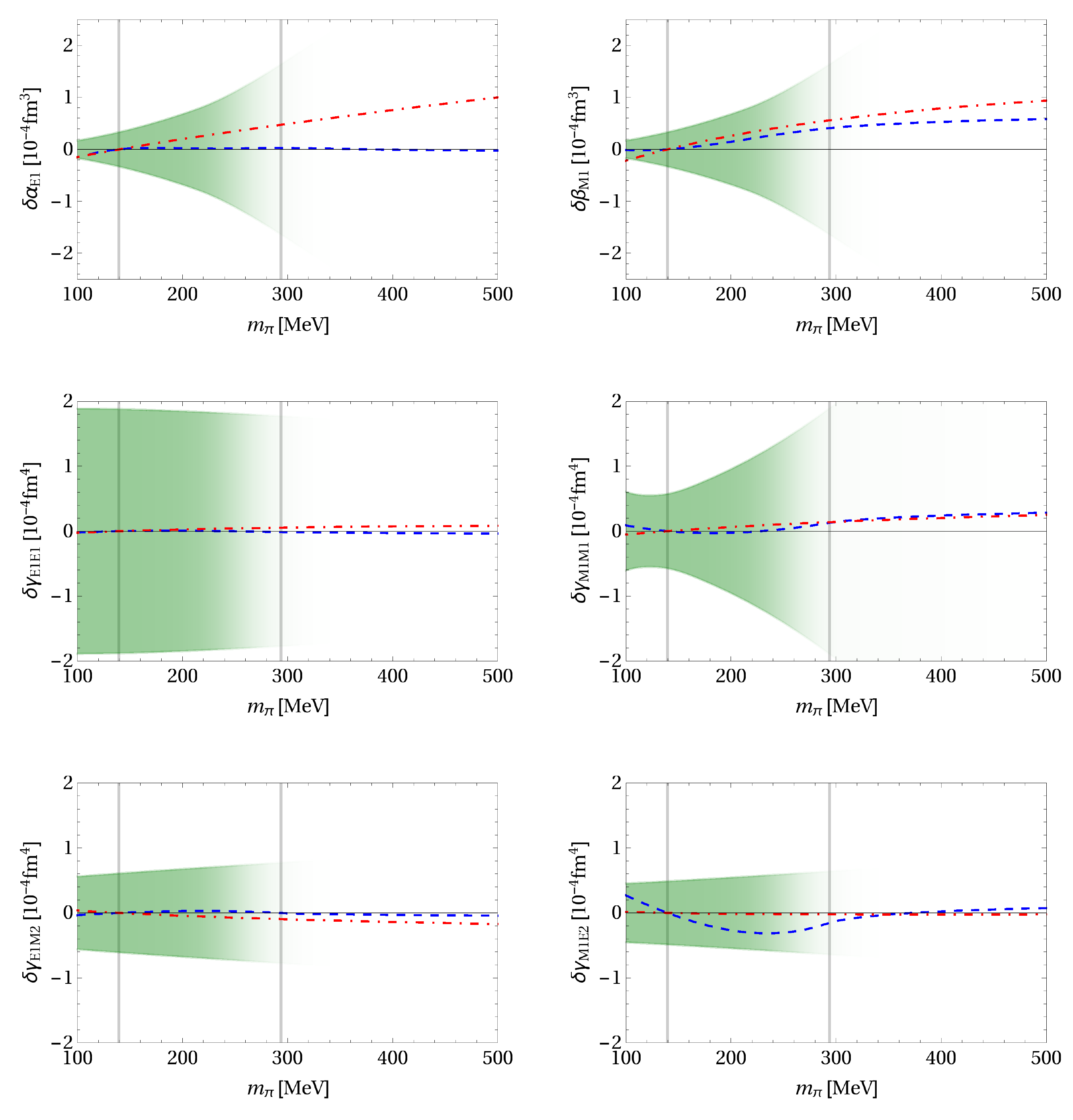}
    \caption{(Colour online) Error bands on theoretical results for
      $\mpi$-dependence of polarisabilities. For greater clarity, the
      $\mpi$-evolution of the central value is not shown. The lines
      demonstrate the added effect of selected next-order corrections to the
      $\mpi$-dependence of the polarisabilities, normalised to zero at
      $\mpiphys$. Blue dashed: $\gamma$N$\Delta$ vertex
      corrections~\cite{McGovern:2012ew}; red dot-dashed: a set of
      ``relativistic" corrections to $\pi\Delta$
      amplitudes~\cite{Lensky:2015awa}. }
    \label{fig:higherorders}
  \end{center}
\end{figure}

Figure~\ref{fig:higherorders} shows the evolution of the error bands for all
six polarisabilities, but \emph{not} the evolution of central values with
$\mpi$, the more clearly to display the behaviour of the uncertainties. As
corroborating evidence for this error assessment, we proceed as in
Sect.~\ref{sec:someho} to take advantage of the fact that three classes of
(isoscalar) pion-mass-dependent higher-order corrections are known. However,
only the first two display $\mpi$-dependence: vertex corrections (blue dashed
line)~\cite{McGovern:2012ew}, and the inclusion of (some) $1/\MN$ corrections
in $\pi \Delta$ loops via a covariant calculation (red dot-dashed
line)~\cite{Lensky:2015awa}. Both effects are much smaller than our $68\%$
uncertainty bands for all $6$ polarisabilities, even in regime (ii), providing
a potential hint that the errors may be over-estimated.  We can safely
conclude that our analysis of uncertainties passes reasonable checks both at
the physical point (Sect.~\ref{sec:values}) and in regard to the $\mpi$
dependence.

\subsection{Theoretical Uncertainties Well Beyond the Physical Pion Mass}
\label{sec:errorregimeii}

For lattice calculations with $\mpi$ markedly larger than $\mpiphys$ the
regime (i) power counting employed in Sect.~\ref{sec:errorregimei} is no
longer appropriate, as discussed in Sect.~\ref{sec:regimes}.  For
$\mpi\sim\DeltaM$ the scales $P(\mpi)$ and $\epsilon$ of
eq.~\eqref{eq:expparams} are of the same order, so there is no suppression of
graphs according to the numbers of Delta propagators in them. Leading $\pi$N
and $\pi\Delta$ loops and Delta-pole graphs are all on the same footing,
contributing to the scalar polarisabilities at $\calO (e^2 \epsilon^{-1})$ and
to the spin polarisabilities at $\calO (e^2 \epsilon^{-2})$. The subleading
$\pi$N loops included above are suppressed by one power of $\epsilon$ but do
not constitute a complete set of contributions at this order. There are, for
instance, Delta-pole graphs with $\pi$N and $\pi\Delta$ loop corrections to
the magnetic and electric $\gamma$N$\Delta$ vertices, and graphs like those of
Fig.~\ref{fig:corrpiN} with intermediate Delta propagators replacing one or
more nucleon propagators. Some, but not all, of the latter are included in a
covariant calculation~\cite{Lensky:2015awa}. Thus, in regime (ii) our
calculation is only complete at leading order; there are omitted effects
already at relative order $\epsilon$.  

Though we have not shown the results here, we did carry out a regime-(ii)
error analysis. We found that for $\mpi\sim \DeltaM$ the bands we generated
were in broad agreement with the regime-(i) bands continued into this
region. This is unsurprising in view of the fact that regime (i) and regime
(ii) are not clearly separated; transition from one to the other is gradual.

In any case, it seems that \ChiEFT does not adequately describe the behaviour
of many observables seen in lattice computations for larger values of
$\mpi$~\cite{McGovern:2006fm,Djukanovic:2006xc,Schindler:2007dr}.  For
example, computations show the nucleon mass rising linearly with $\mpi$ beyond
$300\;\MeV$, not displaying the more complex dependence involving higher
powers predicted by \ChiEFT~\cite{WalkerLoud:2008bp,
  Walker-Loud:2013yua}. Similar difficulties for \ChiEFT are seen in other
observables, too~\cite{Bernard:2006te,Beane:2015}.

Given these issues, we will show predictions for \ChiEFT in regime (ii), but
allow the regime-(i) error bars to fade away beyond $\mpi \approx 250\;\MeV$,
i.e.~as they become unreliable, and disappear altogether beyond about
$350\;\MeV$.  At such pion masses, linear extrapolations in $\mpi$---as used
in Refs.~\cite{Walker-Loud:2013yua,Beane:2015} and earlier studies---may
describe lattice QCD results, but they are not justified within \ChiEFT, and
are more-or-less uncontrolled. We thus refer to results in this regime not as
``chiral predictions'' but as ``chiral curves'' and speculate as to why such
extrapolations may be successful at the end of Sect.~\ref{sec:lattice}.

\subsection{Results: Pion-Mass Dependence of Polarisabilities}
\label{sec:results}

\begin{figure}[!htbp]
\begin{center}
     \includegraphics[width=1.0\textwidth]{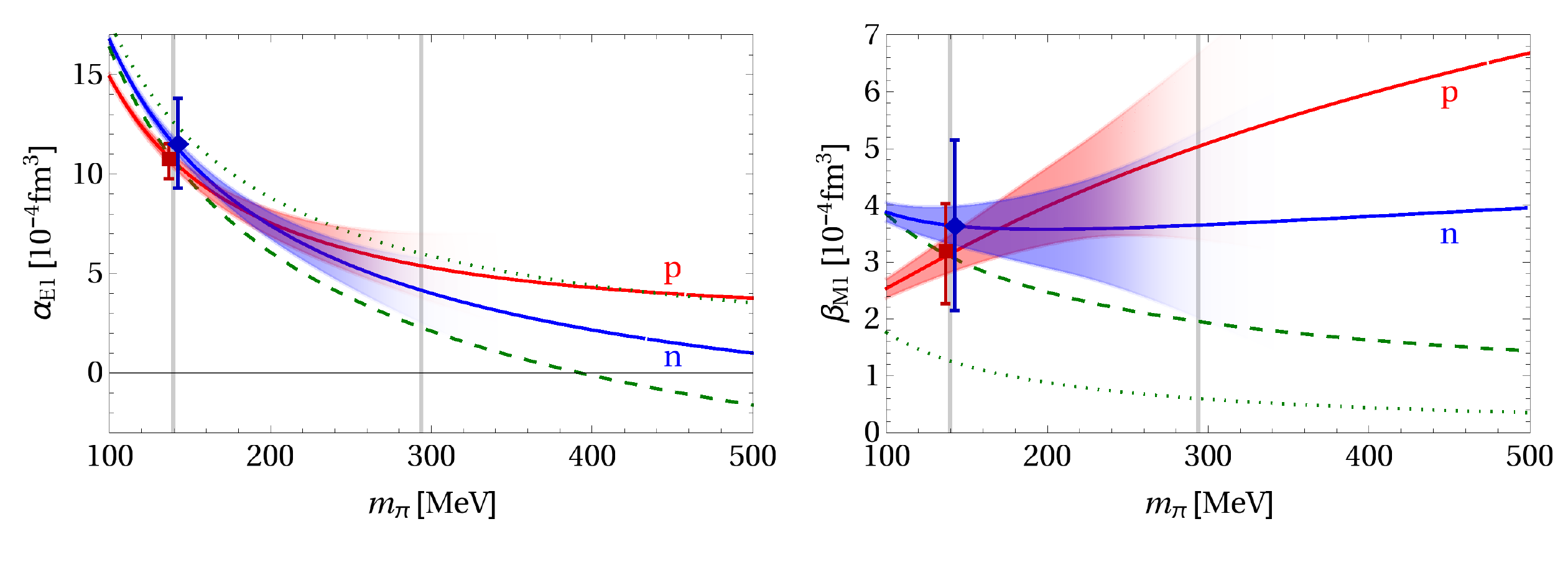}

     \includegraphics[width=1.0\textwidth]{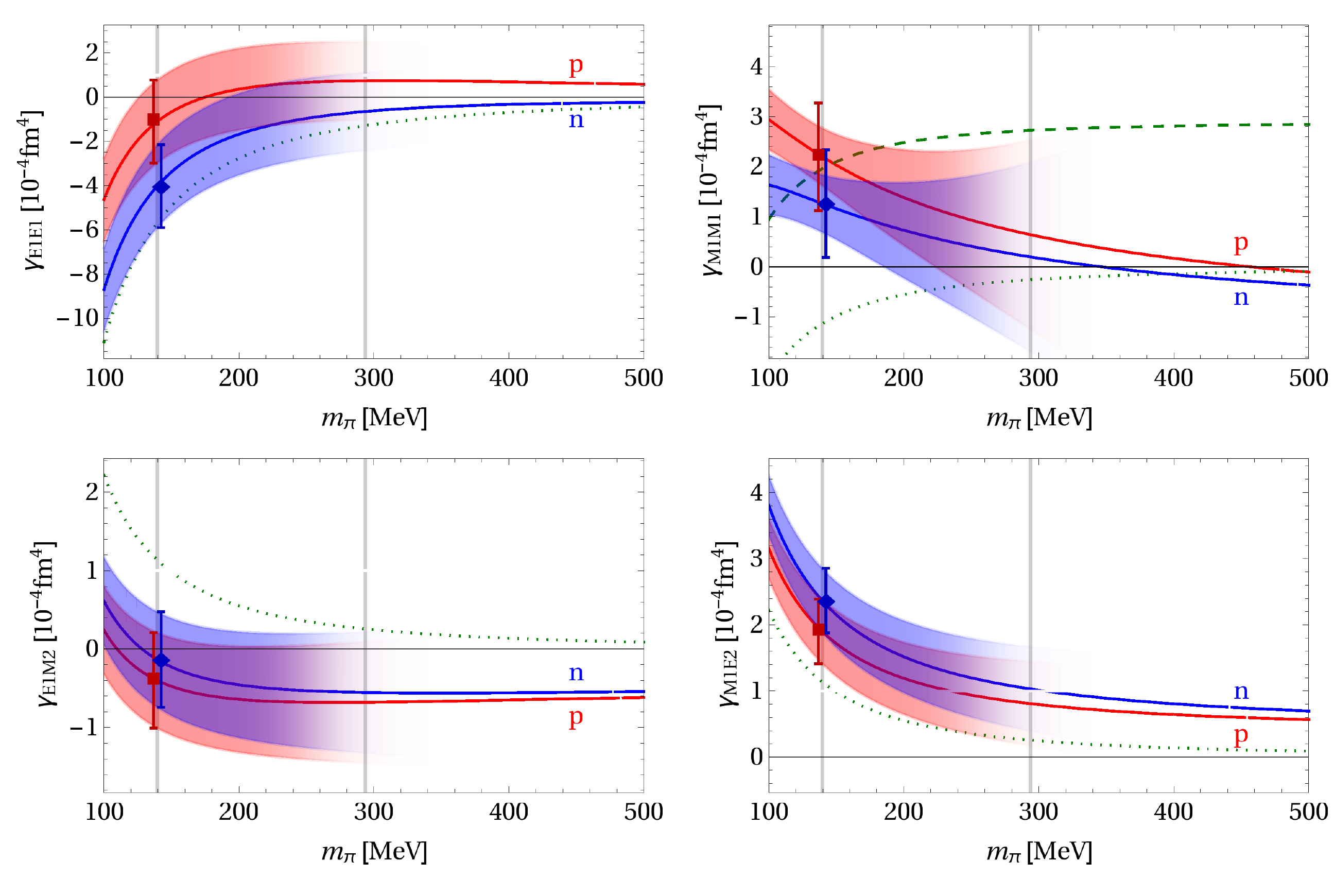}
     \caption{(Colour online) Predicted $\mpi$-dependence of the dipole
       polarisabilities in \ChiEFT. The full result is represented by solid
       lines and labelled for the respective nucleon: proton results are
       coloured red with red corridors of \ChiEFT uncertainties and symbol
       $\protect\textcolor{red}{\blacksquare}$ at $\mpiphys$ (slightly offset
       to smaller $\mpi$ for better visibility); neutron results are blue with
       blue corridors and
       \protect\rotatebox{45}{$\protect\textcolor{blue}{\blacksquare}$} at
       $\mpiphys$ (slightly offset to larger $\mpi$). Error bars at the
       physical point add statistical, theory and Baldin-sum-rule errors
       linearly, as applicable. Green lines are isoscalar, with dotted:
       leading $\pi$N contributions; dashed (for $\alphae$, $\betam$ and
       $\gammamm$ only): Delta pole and leading $\pi\Delta$ pieces added,
       plus isoscalar LECs for $\alphae$, $\betam$. The vertical lines show
       the position of the physical pion mass and the Delta-nucleon mass
       difference. See text for additional details.}
\label{fig:allpols}
\end{center}
\end{figure}

\begin{figure}[!htb]
\begin{center}
  \includegraphics[width=\linewidth]{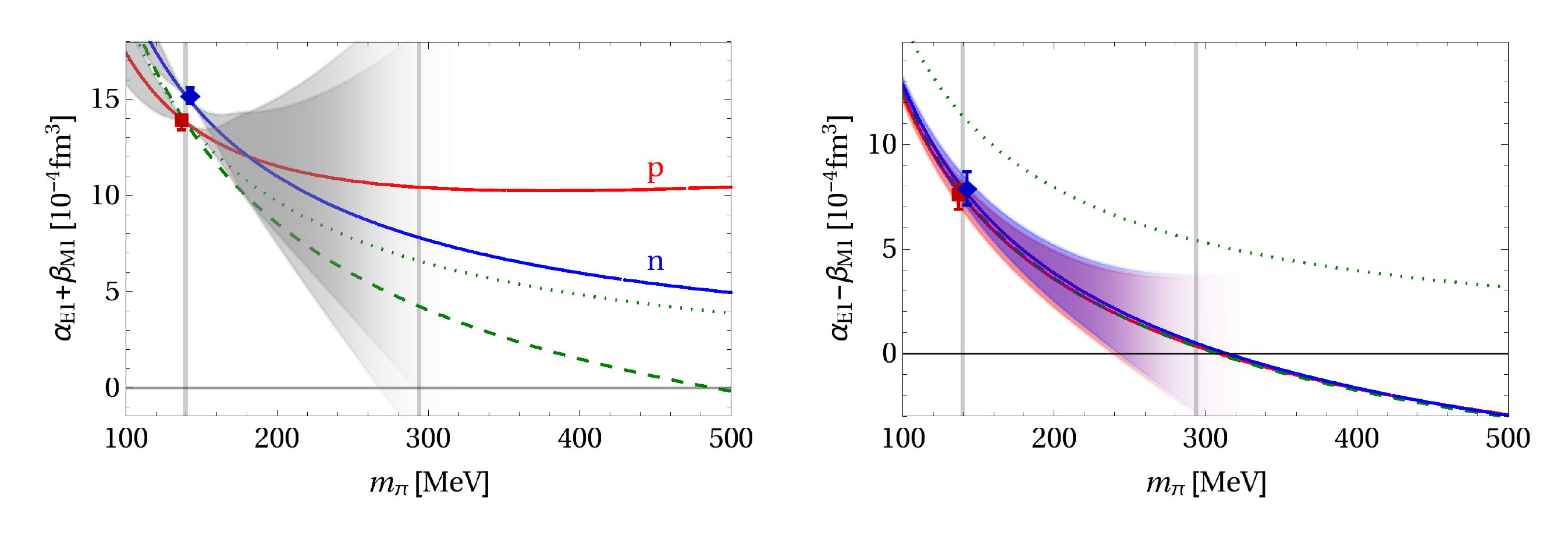}
\\[-3.2ex]
  \includegraphics[width=\linewidth]{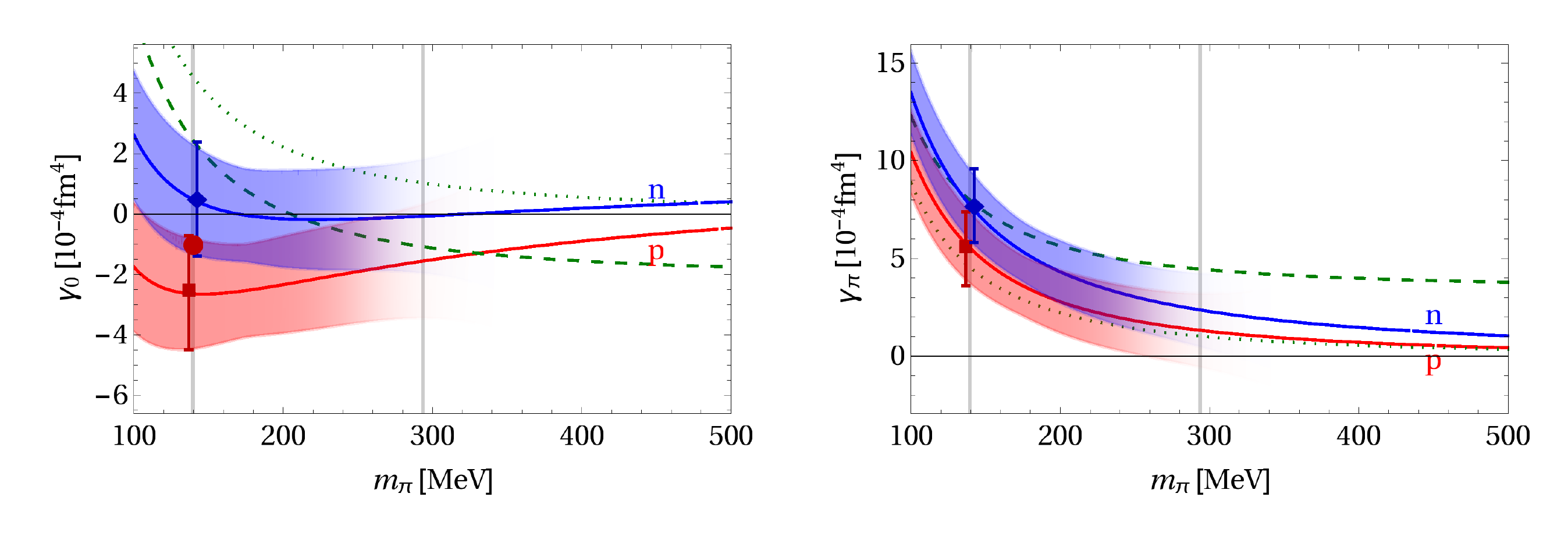}
  \caption{(Colour online) Pion-mass dependence of the Baldin sum rule (top
    left), $\alphae-\betam$ (top right), $\gammazero$ (bottom left) and
    $\gammapi$ (bottom right). For $\alphae+\betam$, error bars at $\mpiphys$
    only reflect the uncertainties of the Baldin sum rules~\cite{Olmos:2001,
      Levchuk:1999zy}; in \ChiEFT, such errors are input and are separate from
    the intrinsic \ChiEFT errors. The \ChiEFT uncertainty corridors of
    $\alphae+\betam$ are coloured differently to indicate that they are not
    used to derive corridors for $\alphae $ or $\betam$.  The experimental
    value of $\gammazerop$ is indicated by the symbol
    $\protect\textcolor{red}{\bullet}$~\cite{Ahrens:2001qt, Dutz}; this
    result's uncertainty is smaller than the size of the symbol. Colour coding
    as in Fig.~\ref{fig:allpols}. See text for further details.}
\label{fig:constraints}
\end{center}
\end{figure}

The solid lines in Figs.~\ref{fig:allpols} and~\ref{fig:constraints} are
\ChiEFT predictions for the pion-mass dependence of the dipole
polarisabilities, together with their theoretical uncertainties, computed as
specified in Sects.~\ref{sec:errorregimei} and~\ref{sec:errorregimeii}. These
are complete at order $\delta^2$ relative to LO for all polarisabilities, and
include three nonzero orders for $\alphae$, $\betam$, $\gammamm$, $\gammazero$
and $\gammapi$, and two nonzero orders for the other polarisabilities. Proton
results are colour-coded in red, neutron ones in blue, and isoscalar ones in
green. The leading $\pi$N loops (green dotted) constitute LO in regime
(i). The green dashed curves represent the complete isoscalar result at order
$\delta^1$ relative to LO in regime (i), as detailed in
Sect.~\ref{sec:Finding-Theory-Uncertainties} and
eq.~\eqref{eq:scalar-pol-values}.  No new contribution arises at this order
for $\gammaee$, $\gammaem$ and $\gammame$, and so this curve is absent
there. In regime (ii), the solid and dashed curves are both only complete to
LO, while the dotted curve is not even the full LO isovector result.

The symbols at the physical point are the \ChiEFT predictions or, for
$\alphae$, $\betam$ and $\gammamm$, fits. Their error bars are found by adding
the theory and---where applicable---statistical plus Baldin-sum-rule errors
linearly, as discussed at the beginning of
Sect.~\ref{sec:extrapolations}. When this total error exceeds the width of the
band at the physical point, the entire band can be floated up or down within
that difference.

We see that the higher-order graphs have only a modest effect on the running
of $\alphae-\betam$, $\gammaee$, $\gammaem$ and $\gammame$ with $\mpi$, but a
major effect in the case of $\gammamm$, $\betam$, $\gammazero$,
$\alphae+\betam$, and, to some degree, $\gammapi$.  In addition,
$\alphae-\betam$ is almost purely isoscalar. At the physical point, this is a
simple consequence of the fact that the fitted proton and neutron values are
very similar, cf.~eq.~\eqref{eq:scalar-pol-values}. Beyond that, its isovector
component grows only logarithmically with $\mpi$ and with a small pre-factor;
see eq.~\eqref{eq:abpiN-corr}.  As already noted, we find it impossible to
assign credible errors to $\alphae+\betam$, because of its poor
convergence. The error bands on $\alphae$ and $\betam$ in
Fig.~\ref{fig:allpols} are therefore simply half those of $\alphae-\betam$,
and are anti-correlated. The uncertainties in the scalar polarisabilities then
appear quite small relative to their magnitudes in both regimes (i) and (ii),
while the uncertainties of the spin polarisabilities are comparable to their
sizes.
\begin{figure}[!tb]
\begin{center}
\includegraphics[width=1.0\textwidth]{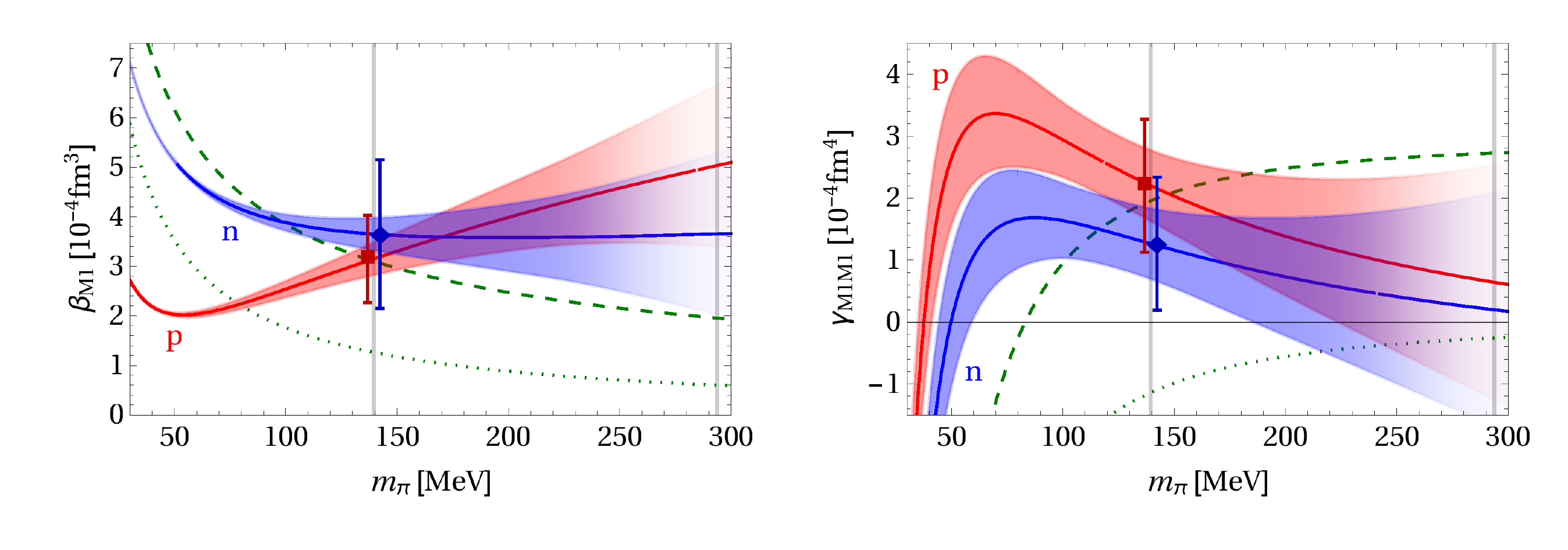}
\caption{(Colour online) Pion-mass dependence of $\betam$ and
  $\gammamm$ in regime (i). Legend as in Fig.~\ref{fig:allpols}. }
\label{fig:zoomin}
\end{center}
\end{figure}

As can be seen in Fig~\ref{fig:allpols}, in most cases the sub-leading $\pi$N
loops do not have a major effect on the trend of the polarisabilities.
However for $\betam$ and $\gammamm$, shown in the chirally relevant
$\mpi$-range in Fig.~\ref{fig:zoomin}, they change functions which are
monotonically decreasing (resp.~increasing) with $\mpi$ at low orders into
ones that are increasing (resp.~decreasing) at $\mpiphys$.  Indeed, both these
polarisabilities could be regarded as somewhat fine-tuned at the physical pion
mass, at least compared to their value at an arbitrary $\mpi < \mpiphys$. This
is interesting in light of the well-known puzzle that the magnetic
polarisability has a physical value which is much smaller than predicted by
the strong paramagnetic effects of the Delta. That higher-order terms in the
$\mpi/\MN$ expansion could change the lower-order trend in $\betam$ was
pointed out in Ref.~\cite{Lensky:2009uv}. At $\mpiphys$, pion-loop effects
cancel against Delta-pole excitations to render the magnetic polarisability
much smaller than either, and produce $\betamp \approx \betamn$. Below the
physical point, the isovector $N\pi$ corrections of eqs.~\eqref{eq:abpiN-corr}
and \eqref{eq:gpiN-corr} gain increasing statistical significance, destroying
this cancellation for both proton and neutron; $\betamn$ quickly approaches
its ``natural'' size in this low-$\mpi$ region, but flattens out once
$\mpi>\mpiphys$, while $\betamp$ changes from a decreasing to an increasing
function around $\mpi=50$ MeV, and after that has a trend that is dictated by
the sub-leading (isovector) $\pi$N loops.

A similar pattern is observed in $\gammamm$, whose leading dependence with
$1/\mpi^2$ dictates a much more prominent ``turn-over'' for $\mpi < \mpiphys$.
In this case the isovector component is smaller than for $\betam$, but it is
still statistically significant.  The fact that $\gammammp$ is somewhat
smaller at the physical point than its generic size for $\mpi < \mpiphys$ may
be related to the necessity to use $\gammamm$ as a parameter in our recent
extraction of the proton's scalar polarisabilities in \ChiEFT,
cf.~Sect.~\ref{sec:CTs}, even though the corresponding LEC only enters one
order higher~\cite{McGovern:2012ew}.

These issues merit further study. For the shape and physical-point value of
these quantities to be markedly affected by contributions of still higher
order, our error corridors would have to underestimate those effects. Our
examination of select higher-order effects in Sect.~\ref{sec:errorregimei}
showed no such problems. Lattice computations closer to the chiral limit would
provide excellent tests, but are numerically quite challenging.

\subsection{Isovector Magnetic Polarisability and the Anthropic Principle}
\label{sec:isovector}

Isovector contributions of sub-leading $\pi$N loops enter at order $\delta^2$
relative to LO for all polarisabilities, and the LECs of the scalar
polarisabilities at the same order have isovector parts as well; see
eq.~\eqref{eq:scalarLEC}. One thus expects that $\alphaev$ and $\betamv$ are
about $20\%$ of $\alphaes$ and $\betams$, respectively. (Recall that we define
$\xi^\text{(s,v)}=\half(\xi^\text{(p)}\pm\xi^\text {(n)})$.)  However,
eq.~\eqref{eq:scalar-pol-values} implies that these isovector combinations are
zero within present uncertainties at the physical pion mass.  This may signal
another instance of fine-tuning between loops and short-distance physics at
the physical point.

This can be quantified via the variation of the isovector polarisabilities
with $\mpi$ (or equivalently $m_\mathrm{q}$).  At this order, the fitted LECs
are $\mpi$-independent, so the relevant rate of change is determined
completely by long-distance physics associated with the subleading $\pi$N
loops of eq.~\eqref{eq:abpiN-corr}:
\begin{equation}
  \label{eq:lnderivative}
  \left.\frac{\dd\betamv}{\dd\ln m_\mathrm{q}}\right|_{\mpiphys}
  =0.65\pm0.4\;\;,\qquad
 \left.\frac{\dd\alphaev}{\dd\ln m_\mathrm{q}}\right|_{\mpiphys}=0.7 \pm 0.4\;\;,
\end{equation}
in $10^{-4}\;\fm^3$, and with a Bayesian estimate of the truncation error.
Therefore, both $\alphaev$ and $\betamv$ vary strongly away from
$\mpiphys$. Indeed, we have already seen in Figs.~\ref{fig:allpols}
and~\ref{fig:zoomin} that the similarity of proton and neutron
polarisabilities disappears for $\mpi\ne\mpiphys$.  While the isovector
component of $\alphae - \betam$ remains small at all $\mpi$ (see
Fig.~\ref{fig:constraints}), the
degeneracy of $\betamp$ and $\betamn$ at the physical point does seem to be
something of an accident. Lattice results at $\mpi=806\;\MeV$ corroborate this
for $\betam$~\cite{Chang:2015qxa}; see Sect.~\ref{sec:lattice} for further
discussion.

The Cottingham Sum rule relates the Compton scattering amplitude to the
electromagnetic part of the proton-neutron self-energy difference,
$\delta\MN^\mathrm{em} \equiv
M_\mathrm{p}^\mathrm{em}-M_\mathrm{n}^\mathrm{em}$~\cite{WalkerLoud:2012bg,
  WalkerLoud:2012en, Erben:2014hza,Thomas:2014dxa,Gasser:2015dwa}, leading us
to speculate about a potential rationale for this apparent coincidence. Though
the topic is not without controversy and there is some scale-dependence in
assigning strong and electromagnetic self-energy differences, there is broad
agreement that $\delta\MN^\mathrm{em}$ and$\betamv$ are connected, since the
latter is related to the value of a component of the integrand in the
Cottingham sum rule at $q^2=0$~\cite{WalkerLoud:2012bg,Erben:2014hza,
  Gasser:2015dwa}. The strength of this connection depends, though, on
assumptions about the integrand, and is hotly debated at present.

According to the analysis of Ref.~\cite{WalkerLoud:2012bg}, at the physical
point the principal contributions to $\delta\MN^\mathrm{em}$ are an elastic
piece of $[0.77\pm0.03]\;\MeV$, and an inelastic piece which is dominated by a
term proportional to $\betamv$:
\begin{equation} 
\label{eq:linkage}
  \delta\MN^{\beta}(\mpi)= -A \,\betamv(\mpi)\;\;.
\end{equation}
The size of $A$ can be obtained from Ref.~\cite{WalkerLoud:2012bg}'s value of
$\delta\MN^\beta\approx0.5\;\MeV$ for $\betamv=-0.5$.
Both the elastic and inelastic part of $\delta \MN^\mathrm{em}$ involve
integrals over form factors, well-known for the elastic contribution and
estimated for the inelastic one.  Assuming the pertinent scale in these form
factors is associated with non-chiral physics, the variation of $\delta
\MN^\mathrm{em}$ with quark mass will come mainly from the magnetic moment in
the elastic term, and from $\betam$ in the inelastic term. Lattice QCD shows,
though, that nucleon magnetic moments are rather insensitive to the quark
mass~\cite{Beane:2014ora}, so we are left with $\delta\MN^{\beta}$ as the
dominant source of variation of $\delta\MN^\mathrm{em}$ with quark mass.

If we assume $A$ of eq.~\eqref{eq:linkage} is $m_\pi$ independent, we can
estimate the variation of $\delta\MN^\beta$ as
\begin{equation}
  \left.\frac{\dd\delta\MN^{\beta}(\mpi)}{\dd\ln  
      m_\mathrm{q}}\right|_{\mpiphys}=-0.65\;\MeV\;\;.
      \label{eq:estimate}
\end{equation}
Here, we do not give uncertainties, since we cannot quantify them on some of
the assumptions being made. This value is not negligible relative to the
quark-mass variation in Ref.~\cite{Bedaque:2010hr}:
$\left.\frac{\dd\delta\MN^\text{strong}}{\dd\ln
    m_\mathrm{q}}\right|_{\mpiphys}\approx-2.1\;\MeV$ (obtained under the
assumption that $m_\mathrm{u}/m_\mathrm{d}$ remains constant). Of course, the
slope of $\delta\MN^{\beta}$ could be smaller than our estimate, or somehow
cancelled by other effects in $\delta\MN^{\rm em}$. But the estimate
\eqref{eq:estimate} makes it plausible that---contrary to what was assumed
heretofore in many works, such as Ref.~\cite{Bedaque:2010hr}---the variation
of $\delta\MN^\mathrm{em}$ is not negligible in comparison to that of the
strong part.

If $\delta\MN^\beta$ does indeed produce the largest variation of $\delta
\MN^\mathrm{em}$ with $m_\mathrm{q}$, our estimate suggests that the
quark-mass dependence of the proton-neutron mass splitting may be
significantly enhanced---or indeed reduced for $\betamp>\betamn$, which is
within today's allowed range.  As a consequence, the neutron life-time would
then either be substantially shortened as $\mpi$ increases (if
$\betamp<\betamn$), or as $\mpi$ decreases (if $\betamp>\betamn$).  A neutron
that is too short-lived to allow Big-Bang Nucleosynthesis to proceed to
${}^4$He presumably makes carbon-based life impossible.  This putative
connection between a small $\betamv$ and the Anthropic Principle deserves
additional investigation.

On a more prosaic level, the connection to $\delta \MN^\beta$ was used by
Thomas et al.~to extract values for $\betamv$ from the RBC lattice results for
the electromagnetic self-energy of the nucleon~\cite{Blum:2010ym}. They
deduced small negative values (between $-0.5$ and $0$) at four pion masses
between $279$ and $683\;\MeV$~\cite{Thomas:2014dxa}. Neither our results for
$\betamv$ nor the lattice computations by NPLQCD support this
finding~\cite{Chang:2015qxa}; see Sect.~\ref{sec:lattice}. We point out here
that the analysis of Ref.~\cite{Thomas:2014dxa} assumes that other
contributions to $\delta \MN^{\rm em}$ are completely negligible.  Comparisons
of direct measurements of polarisabilities in lattice QCD with our analysis
are more satisfying, and it is to these that we now turn.

\section{Comparison with Lattice Computations}
\setcounter{equation}{0}
\label{sec:lattice}

\begin{figure}[!htb]
\begin{center}
     \includegraphics[width=\textwidth]{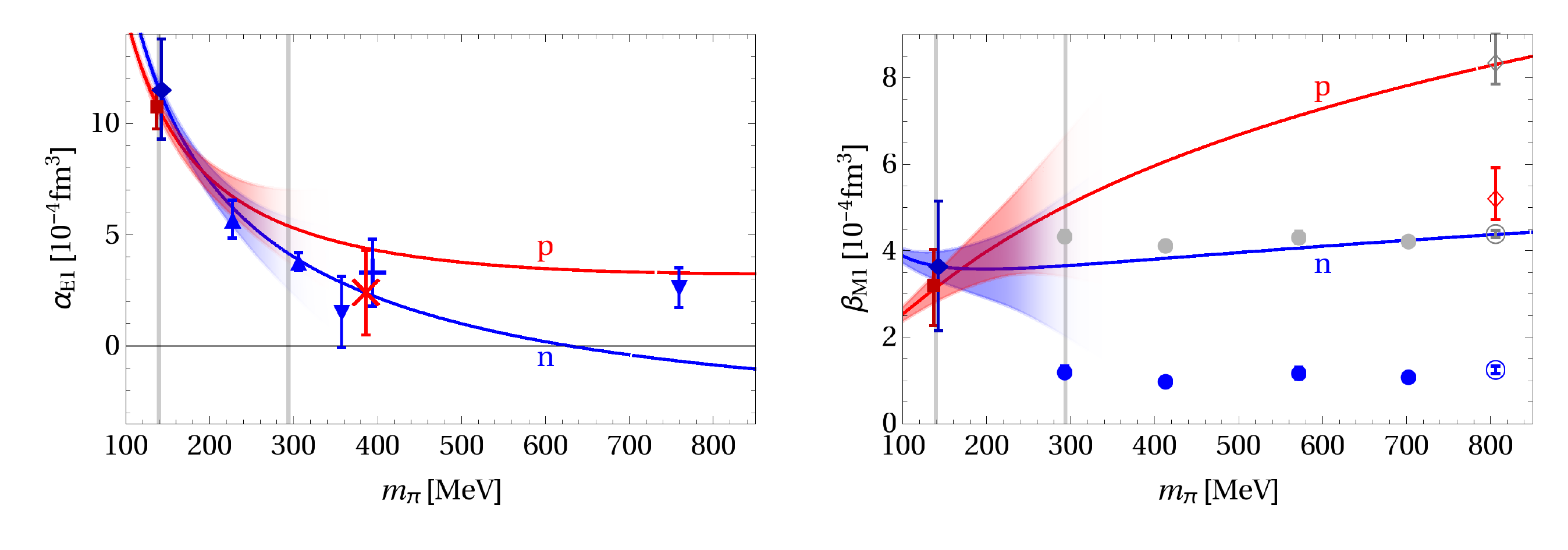}\\[-0ex]
     \includegraphics[width=0.475\textwidth]{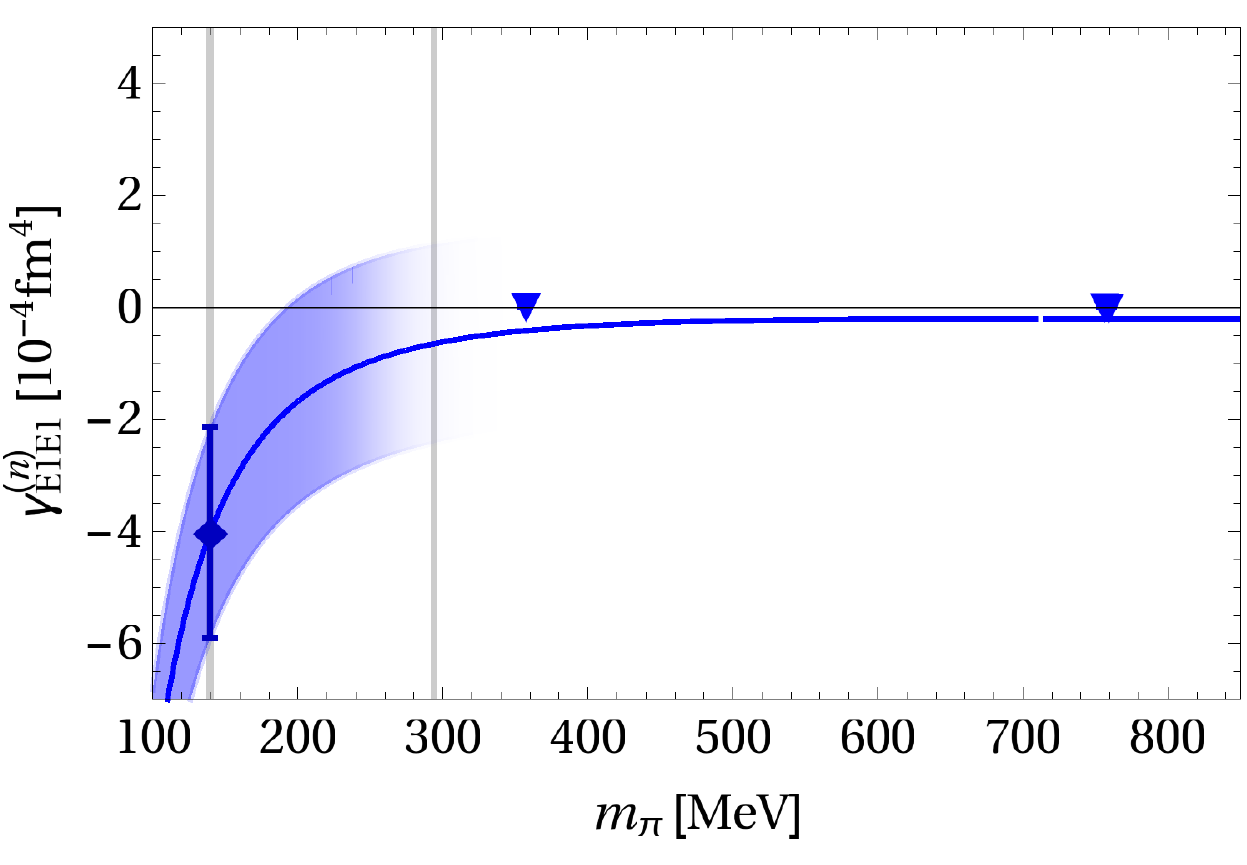}
     \caption{(Colour online) Comparison to lattice-QCD computations.  Lattice
       computations of $\alphae$ (top left):
       $\protect\textcolor{blue}{\blacktriangle}$ (neutron) Lujan et
       al.~\cite{Lujan:2014qga}; $\protect\textcolor{red}{\times}$ (proton)
       and $\protect\textcolor{blue}{+}$ (neutron) Detmold et
       al.~\cite{Detmold:2010ts} (slight $\mpi$-offset for better visibility);
       $\protect\textcolor{blue}{\blacktriangledown}$ (neutron)
       Engelhardt/LHPC~\cite{Engelhardt:2007ub,
         Engelhardt:2010tm}.  
Lattice computations of $\betam$ (top right): $\protect\textcolor{blue}{\bullet}$ (neutron) Hall et al.~\cite{Primer:2013pva, Hall:2013dva}; \protect\rotatebox{45}{$\protect\textcolor{red}{\scriptscriptstyle\square}$} (proton) and $\protect\textcolor{blue}{\circ}$ (neutron) NPLQCD~\cite{Chang:2015qxa}.  Gray ``ghost points'' found by shifting all lattice results by $+3\times10^{-4}\;\fm^3$.
Lattice computation of $\gammaeen$ (bottom):
$\protect\textcolor{blue}{\blacktriangledown}$ (neutron) Engelhardt et
al.~\cite{Engelhardt:2011qq, Engelhardt:2015}, see text for qualifier. For
Refs.~\cite{Primer:2013pva, Hall:2013dva, Engelhardt:2011qq, Engelhardt:2015},
the reported lattice errors are smaller than our symbol sizes. Further
notation as in Fig.~\ref{fig:allpols}, including
$\protect\textcolor{red}{\blacksquare}$ for proton values and
\protect\rotatebox{45}{$\protect\textcolor{blue}{\blacksquare}$} for neutron
ones at $\mpiphys$.}
\label{fig:lattice}
\end{center}
\end{figure}

In Fig.~\ref{fig:lattice}, we compare our findings to emerging lattice-QCD
computations of dipole polarisabilities. We do not report calculations without
sea quarks---i.e.~those that set the fermion determinant to one---and we have
selected only references for pion masses up to about $850\;\MeV$, with values
that were either extrapolated to infinite volume and infinitesimal lattice
spacing, or for which the authors estimated such effects to be irrelevant at
present accuracies; cf.~Refs.~\cite{Detmold:2006vu, Tiburzi:2014zva}. To our
knowledge, the work by Engelhardt/LHPC~\cite{Engelhardt:2007ub,
  Engelhardt:2010tm} on $\alphaen$ is the only extant calculation which meets
these criteria and also accounts for the charges of the sea quarks
themselves. At present, all other computations use uncharged sea quarks whose
mass is identical to that of the valence quarks.  Several efforts to include
charged sea-quark effects are ongoing~\cite{Freeman:2014kka,
  Engelhardt:2007ub, Engelhardt:2010tm}. We report lattice uncertainties as
stated in the sources; a thorough appraisal of the lattice computations is not
our goal. Note that an analysis of the consistency of lattice computations
with our $\mpi$-dependent predictions for polarisabilities cannot proceed by
simple ``standard-deviation counting", because the uncertainties in the shape
of $\xi(\mpi)$, and thus also the theory errors at different pion masses, are
highly correlated. Ref.~\cite{Stump:2001gu} derives a modified $\chi^2$, whose
use would be one way to account for such systematic errors.

We find that the \ChiEFT prediction for $\alphaen$ agrees well with the
available computations, of Lujan et al.~\cite{Lujan:2014qga}, Detmold et
al.~\cite{Detmold:2010ts}, and Engelhardt/LHPC~\cite{Engelhardt:2007ub,
  Engelhardt:2010tm}. For the proton, only the result of Detmold et al.~for
$\alphae$ at $\mpi\approx400\;\MeV$ meets our selection criteria; it is quite
compatible with the chiral curve. In regime (i) and (ii) these are
statistically rigorous statements, since our error corridors provide estimates
of higher-order effects---albeit with decreasing reliability as $\mpi$
increases.  The agreement for electric polarisabilities is so good that we
feel we must stress that our results are not fits to lattice computations; the
physical value is determined by Compton-scattering data on the proton and
deuteron, and the pion-mass dependence exclusively by chiral dynamics.

For the magnetic polarisability of the neutron, $\betamn$, we cannot reproduce
the magnitude reported by the Adelaide group~\cite{Primer:2013pva,
  Hall:2013dva}.  NPLQCD~\cite{Chang:2015qxa} also reports results for
$\betamn$ and $\betamp$ at $806\;\MeV$; the former agrees well with the
Adelaide results.  The authors of this paper point out that, since this pion
mass corresponds to the flavour SU(3) limit, their result for the isovector
combination $\betamp-\betamn$ is unaffected by their neglect of sea-quark
charges, in contradistinction to the isoscalar one.  Interestingly, their
isovector result agrees nearly exactly with the chiral curve.  This is
illustrated by the grey ``ghost points'' in the top right panel of
Fig.~\ref{fig:lattice}, to which we have added constant isoscalar contribution
of $3\times10^{-4}\;\fm^3$.  This uncannily good agreement---at a pion mass of
$800\;\MeV$, which is certainly outside the radius of convergence of
\ChiEFT---may of course be pure coincidence.  Nevertheless, the lattice result
of a sizeable isovector splitting at $\mpi=806\;\MeV$ seems to support our
finding in Sect.~\ref{sec:isovector} that $\betamp \approx \betamn$ at the
physical point is something of a coincidence. Lattice QCD calculations at
intermediate pion masses will either strengthen or rebut this conclusion. They
can also check if the \ChiEFT prediction of a significant $\alphaev$ for $\mpi
> \mpiphys$ is realised in QCD.

Engelhardt et al.~have also reported the first lattice study of a spin
polarisability~\cite{Engelhardt:2011qq, Engelhardt:2015}. The neutron's
$\gammaee$ shows a minuscule but nonzero signal within the reported
uncertainties. These values are without the subtraction of the pion-pole
contribution in eq.~\eqref{eq:pi-pole}.  However, a calculation with uncharged
sea quarks, like this one, has a number of pathologies. For example, in the
two-flavour variant considered in Ref.~\cite{Detmold:2006vu}, the isovector
``pion-pole'' contribution of the physical particle $\propto\ga\tau_3$ must be
supplemented by a degenerate isoscalar-scalar ghost which couples with an
unknown strength $g_1$ to the nucleon. If $g_1$ has a similar magnitude to
$\ga$, then even the sign of the total ``pion-pole" contribution is
unknown. The total lattice values of $\gammaeen$ are extremely small, so in
the absence of a very fine-tuned cancellation between ``pion-pole'' and
structure contributions, it is likely that both are small. We therefore feel
justified in placing the lattice points---which include both the physical and
pathological pion-pole contributions---on the same plot as our
pion-pole-subtracted curves.

In most cases, lattice groups account for the differences between identifying
polarisabilities as the terms quadratic in the electromagnetic fields, and the
canonical definition via non-pole parts of the Compton amplitudes; see
Refs.~\cite{Lvov:1993fp, Bawin:1996nz, Schumacher:2005an, Lee:2014iha} for
further discussion.  We follow Ref.~\cite{engelprivcomm} in adding the
Dirac-Foldy contribution of $\alpha_{\mathrm{EM}}(\kappan)^2/(4\MN^3)\approx
0.7$ to $\alphaen$ in Ref.~\cite{Engelhardt:2010tm}. Similarly,
$\alpha_{\mathrm{EM}}/(4\MN^3)\approx0.2$ should be subtracted from $\betamp$
at $\mpi=806\;\MeV$ in Ref.~\cite{Chang:2015qxa}, but the effect is well
within the lattice uncertainties~\cite{tiburziprivcomm}.

The surprising agreement between some of our chiral curves and lattice results
at very large pion masses may be accidental. This agreement occurs far outside
the \ChiEFT radius of convergence, and could just be a coincidence. But it is
striking, and so we close this section with a testable speculation as to why
it could be more than an accident.  First we note that, at the order to which
we work, the chiral expansion does not produce positive powers of $\mpi$; our
\ChiEFT result includes at most a logarithmic divergence as $\mpi\to\infty$.
Were we to go further in the expansion, we would encounter the usual problem
of contributions that grow with more and more powers of $\mpi$.  The very mild
dependence on pion mass seen in several lattice observables (see
e.g.~\cite{WalkerLoud:2008bp, Walker-Loud:2013yua}) can then only be
reconciled with \ChiEFT through ever-increasing fine-tuning between terms of
different, higher orders. In the case of the lattice polarisabilities for
$\mpi\gtrsim 400\;\MeV$, such pion-mass-independence is more akin to that
expected in a heavy-constituent-quark model, or in the classical Lorentz model
which considers heavy, charged particles in a harmonic-oscillator potential.
Indeed, this smooth $\mpi$-evolution might be considered generic. Of course,
QCD must provide a smooth interpolation between the chiral and heavy-quark
regimes; cf.~Ref.~\cite{Leinweber:1999ig} for the case of baryon masses.  It
is plausible that results like ours, with weak $\mpi$-dependence at low chiral
orders and large pion masses, provide reasonable extrapolations into the
regime in which \ChiEFT is \emph{a priori} inapplicable.  Such a ``principle
of chiral persistence'' suggests that our chiral extrapolations for the spin
polarisabilities may turn out to match future lattice computations at high
pion masses, too. If such ``chiral persistence" is not a feature of QCD, then
those computations will reveal the agreement seen thus far to indeed be
accidental.

\section{Summary and Conclusions}
\setcounter{equation}{0}
\label{sec:conclusions}

In this paper, we have presented the static scalar and spin dipole
polarisabilities of both the proton and neutron in \ChiEFT as a function of
the pion mass. We have included the leading and sub-leading effects of the
nucleon's pion cloud, together with the leading contributions of the
$\Delta(1232)$ and its pion cloud. We have differentiated between two
pion-mass regimes. Close to the physical $\mpi$, our results are complete at
second order in the small expansion parameter
$\delta\approx\sqrt{\mpi/\Lambda_\chi}\approx \DeltaM/\Lambda_\chi$.  This
corresponds to three non-vanishing orders for the scalar polarisabilities and
two for the spin polarisabilities. For $\mpi\sim\DeltaM$, however, the results are
complete only at leading order since contributions are reordered: leading
$\pi$N, $\Delta(1232)$ and $\pi\Delta$ effects are all of similar size.

A central goal of the paper is to provide reproducible estimates of
uncertainties from within the \ChiEFT framework which are as objective as
feasible. At each pion mass, we have used a recently developed statistical
interpretation of standard order-by-order EFT convergence estimates to derive
$68\%$ degree-of-belief intervals. The resulting probability distributions are
non-Gau\3ian. They are based on several assumptions: the error associated with
the first omitted term in the \ChiEFT series dominates the uncertainty; the
corresponding EFT coefficient is ``natural'' in units of the breakdown scale;
and the size of this first omitted term grows linearly with $\mpi$. The
inclusion of select higher-order effects indicates that for pion masses below
about $250\;\MeV$ our uncertainties are, if anything, overestimated. In fact,
basic physical arguments imply that our truncation-error for $\alphae+\betam$
is markedly too large for $\mpi > \mpiphys$, so we somewhat arbitrarily assign
zero truncation error to it. Truncation errors must be combined with
uncertainties in the input parameters, like the error on $\alphae + \betam$
from the Baldin sum rule. A framework that combines all these errors in a
statistically consistent way is under development~\cite{Furnstahl:2014xsa,
  Wesolowski:2015fqa}. 

At the physical pion mass, the truncation errors of the spin polarisabilities
augment our previously published prediction of the central values, and our
recent fits of the scalar polarisabilities. In all cases, the Bayesian method
provides a rigorous theory error. Our spin polarisabilities agree well, within
their errors, with available extractions from data, and with the predictions
of both dispersion relations and a formulation of \ChiEFT with relativistic
baryons.

For the neutron electric polarisability and $\gammaeen$, agreement between our
\ChiEFT predictions and extant lattice computations at $\mpi \lsim 350$ MeV is
remarkably good.  Beyond this pion mass, there are doubts about the
convergence of \ChiEFT, and the error bars we derived are certainly not
trustworthy. Nevertheless, if we extrapolate the central value of our \ChiEFT
curves into this regime, agreement persists for lattice results on both
$\alphaen$ and $\alphaep$ at $\mpi \approx 400$ MeV---within the uncertainties
on the lattice numbers.  Taking such an extrapolation for the isovector
magnetic polarisability out to $\mpi \approx 800$ MeV yields striking
agreement with the recent results of Ref.~\cite{Chang:2015qxa}.  This is
surprising, given that \ChiEFT is certainly not convergent at such a large
pion mass.  We speculate that such an extrapolation of the \ChiEFT curve does
better than we have any right to expect, because---at the order to which we
work---the pion-mass dependence is tame enough to permit smooth evolution into
the functional dependence of a constituent-quark model.

While most of the present lattice results are at too high a pion mass to be
reliably extrapolated to the physical point, they still corroborate important
aspects of our findings. A cancellation between $\pi$N loops and
short-distance mechanisms encoded in LECs makes the magnetic scalar and spin
polarisabilities small at $\mpiphys$, but this cancellation does not persist
away from the physical point. Similar fine-tuning leads to a physical-world
proton-neutron difference that, for both the scalar electric and magnetic
polarisabilities, is consistent with zero within present uncertainties. Both
of these proton-neutron degeneracies are lifted away from the physical point.
In the near future, lattice calculations could examine the onset of these
cancellations as $\mpi$ is lowered towards, and even below, $250\;\MeV$. We
pointed out that $\betamv$ may have a previously-neglected impact on the
variation of the proton-neutron mass difference with the quark mass.  The
critical role of the neutron lifetime in Big-Bang Nucleosynthesis then
suggests an anthropic argument may explain what otherwise appears to be a
coincidentally small value of $\betamv$.

Finally, we look forward to a next-order calculation of the polarisabilities
in \ChiEFT. This includes subleading effects of the pion cloud around the
Delta. The associated LECs in the \ChiEFT with an explicit $\Delta(1232)$ are,
in principle, constrained by data from other processes, e.g.~$\pi$N scattering
and pion photoproduction. But in practice, their values have non-negligible
uncertainties~\cite{Gellas:1998wx, Pascalutsa:2006up, Yao:2016vbz}. Still,
such an ${\cal O}(e^2 \delta^5)$ calculation could yield more accurate
predictions at the physical point, in particular for spin polarisabilities and
isovector parts. It may also allow a better assessment of the convergence of
the chiral series for $\mpi\sim\DeltaM$, where only leading-order accuracy is
available at present.

\section*{Acknowledgements}

This project was prompted by a request from A.~Alexandru. We are indebted to
him and to M.~Lujan for discussions and encouragement; to M.~Savage for
well-timed advice, crucial inspiration, and sharing findings of the NPLQCD
collaboration prior to publication; and to S.~Beane for discussions on the
behaviour of low-energy quantities outside the chiral regime. We also
gratefully acknowledge correspondence with A.~Alexandru, M.~Engelhardt and
B.~Tiburzi concerning details of their respective lattice computations, and
with M.~Hoferichter and H.~Leutwyler on the impact of $\betamv$ on the
proton-neutron mass difference. Finally, we are grateful to the organisers and
participants of the workshops \textsc{Compton Scattering off Protons and Light
  Nuclei: Pinning Down the Nucleon Polarizabilities} (2013) and
\textsc{Lattice Nuclei, Nuclear Physics and QCD---Bridging the Gap} (2015),
both at the ECT*, Trento (Italy), and \textsc{Bound States and Resonances in
  Effective Field Theories and Lattice QCD Calculations}, Benasque (Spain,
2014) for financial support, and stimulating presentations and
atmosphere. This work was supported in part by UK Science and Technology
Facilities Council grants ST/J000159/1 and ST/L005794/1 (JMcG), by the US
Department of Energy under contracts DE-FG02-93ER-40756 (DRP), as well as
DE-FG02-95ER-40907 and DE-SC0015393 (HWG), and by the Dean's Research Chair
programme of the Columbian College of Arts and Sciences of The George
Washington University (HWG).


\newpage

\appendix
\section{Shapes and Profiles of Corridors of
  Uncertainties}
\setcounter{equation}{0}
\label{app:appendix}

\begin{figure}[!htb]
\begin{center}
\includegraphics[clip=,width=\linewidth]{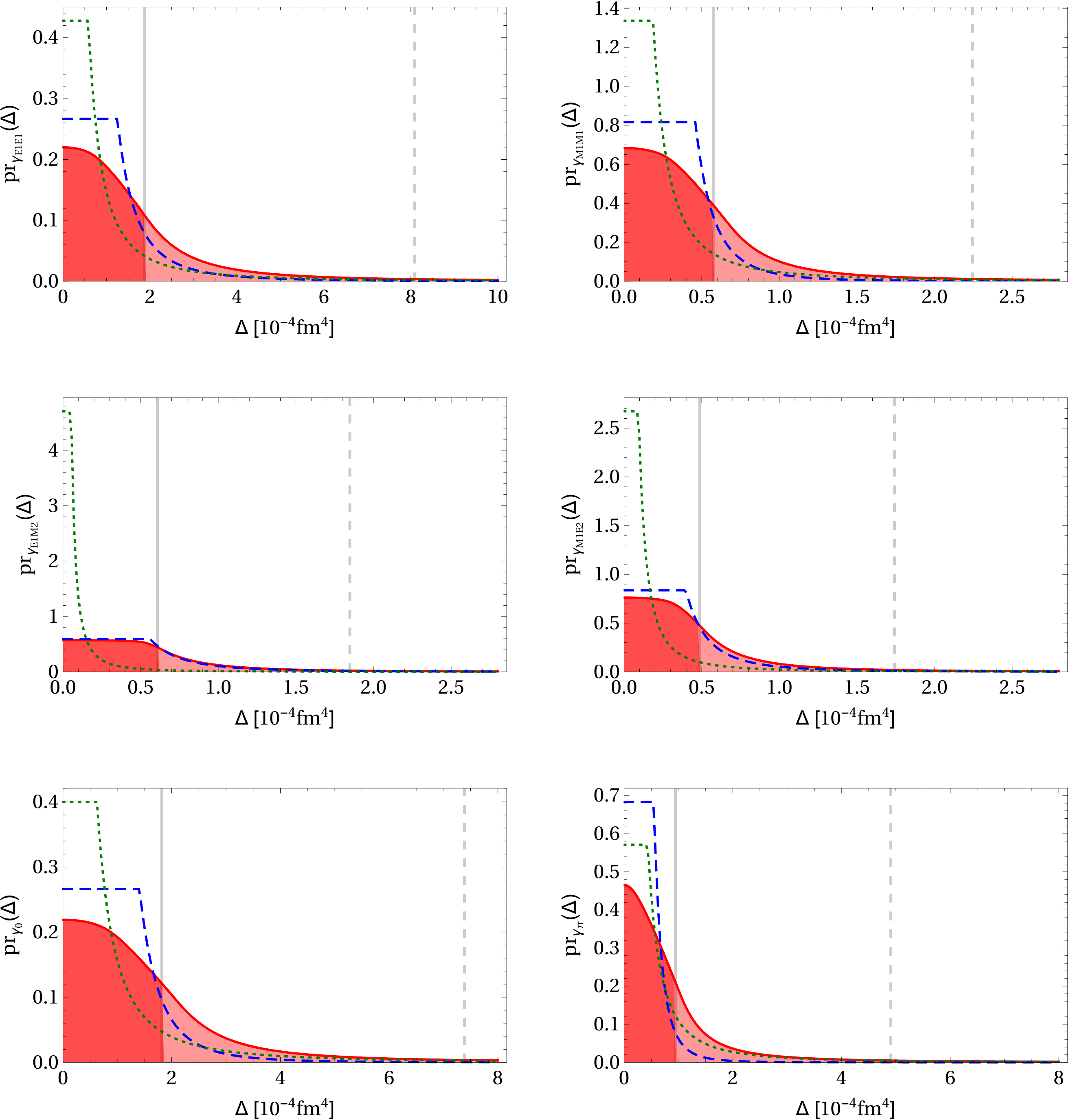}
\caption{(Colour online) Pdfs for the spin polarisabilities at the
  physical pion mass. Notation as in Fig.~\ref{fig:combinedpdfs}.}
\label{fig:spinposterior}
\end{center}
\end{figure}

\begin{figure}[!htb]
\begin{center}
\includegraphics[clip=,width=\linewidth]{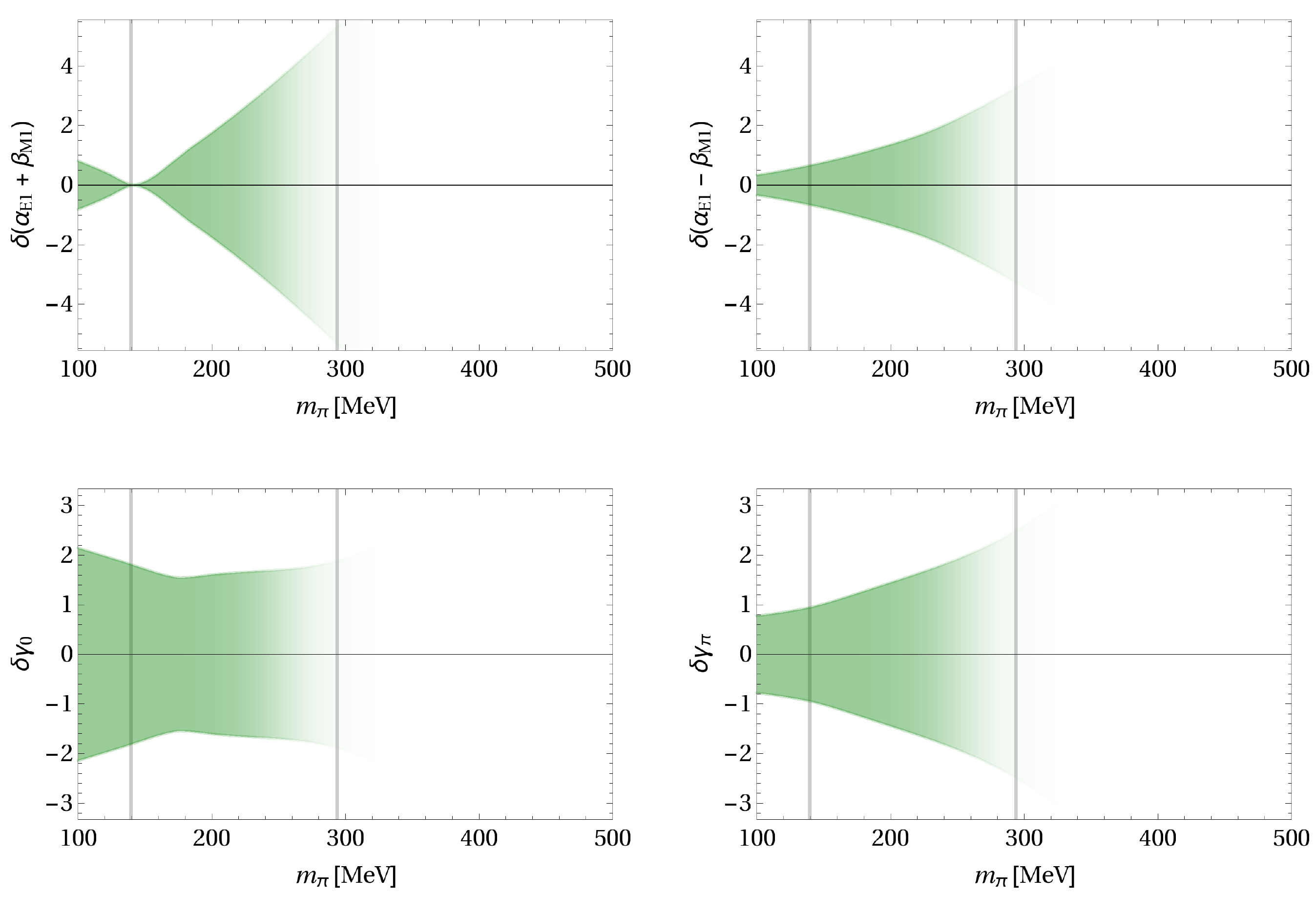}
\caption{(Colour online) Error bands ($68\%$ DoBs) on theoretical results for
  the $\mpi$-dependence of polarisability combinations $\alphae\pm\betam$
  (top) and $\gamma_{0,\pi}$ (bottom). For $\gammapi$, the original result of
  the Bayesian analysis is shown. In Fig.~\ref{fig:constraints}, it is
  substituted by that for $\gammazero$, as detailed in
  Sect.~\ref{sec:Finding-Theory-Uncertainties}.}
\label{fig:higherorders-supplement}
\end{center}
\end{figure}

We show in Fig.~\ref{fig:spinposterior} the pdfs for all polarisabilities at
the physical pion mass, and in Fig.~\ref{fig:higherorders-supplement} the
theoretical error corridors of the results for the $m_\pi$-dependence of
$\alpha_{E1}\pm\beta_{M1}$, $\gamma_0$ and $\gamma_\pi$, as detailed in
Sect.~\ref{sec:Finding-Theory-Uncertainties}.

\begin{thebibliography}{99}

\bibitem{Griesshammer:2012we}
  H.~W.~Grie\3hammer, J.~A.~McGovern, D.~R.~Phillips and G.~Feldman,
  Prog.\ Part.\ Nucl.\ Phys.\  {\bf 67} (2012) 841
  [arXiv:1203.6834 [nucl-th]].

\bibitem{Jenkins:1991ne} E.~E.~Jenkins and A.~V.~Manohar, In \emph{Effective
    field theories of the standard model}, 113 (proceedings Dobogokoe 1991,
  ed.~by U.-G.~Mei\3ner, World Scientific 1992) and Calif.~Univ.~San Diego
  report No.~UCSD-PTH 91-30 (91/10, rec.~Dec.) 26 p.~(201392).

\bibitem{Bernard:1991rq}
  V.~Bernard, N.~Kaiser and U.-G.~Mei\3ner,
  Phys.\ Rev.\ Lett.\  {\bf 67 } (1991)  1515.

\bibitem{Bernard:1995dp}
  V.~Bernard, N.~Kaiser and U.-G.~Mei\3ner,
  Int.\ J.\ Mod.\ Phys.\  E {\bf 4} (1995) 193
  [arXiv:hep-ph/9501384].

\bibitem{Butler:1992ci}
  M.~N.~Butler and M.~J.~Savage,
  Phys.\ Lett.\ B {\bf 294} (1992) 369
  [hep-ph/9209204].

\bibitem{Hemmert:1996xg}
  T.~R.~Hemmert, B.~R.~Holstein and J.~Kambor,
  Phys.\ Lett.\ B {\bf 395} (1997) 89
  [hep-ph/9606456].

\bibitem{Hemmert:1997ye}
  T.~R.~Hemmert, B.~R.~Holstein and J.~Kambor,
  J.\ Phys.\ G {\bf 24} (1998) 1831
  [hep-ph/9712496].

\bibitem{Pascalutsa:2002pi}
  V.~Pascalutsa and D.~R.~Phillips,
  Phys.\ Rev.\  C {\bf 67}  (2003) 055202.
  [nucl-th/0212024].

\bibitem{Hildebrandt:2003fm} 
  R.~P.~Hildebrandt, H.~W.~Grie\3hammer, T.~R.~Hemmert and B.~Pasquini,
  Eur.\ Phys.\ J.\  A {\bf 20} (2004) 293
  [arXiv:nucl-th/0307070].

\bibitem{McGovern:2012ew}
  J.~A.~McGovern, D.~R.~Phillips and H.~W.~Grie\3hammer,
  Eur.\ Phys.\ J.\ A {\bf 49} (2013) 12
  [arXiv:1210.4104 [nucl-th]].

\bibitem{Myers:2014ace} 
  L.~S.~Myers {\it et al.}  [COMPTON@MAX-lab Collaboration],
  Phys.\ Rev.\ Lett.\  {\bf 113} (2014) 262506
  [arXiv:1409.3705 [nucl-ex]].

\bibitem{Weller:2009zz}
  H.~R.~Weller, M.~W.~Ahmed, H.~Gao, W.~Tornow, Y.~K.~Wu, M.~Gai and R.~Miskimen,
  Prog.\ Part.\ Nucl.\ Phys.\  {\bf 62} (2009) 257.

\bibitem{HIGSPAC} \HIGS Programme-Advisory Committee Reports 2009 to 2016,
  with list of approved experiments at  \href{http://www.tunl.duke.edu/higs/experiments/approved/}{www.tunl.duke.edu/higs/experiments/approved/} 

\bibitem{Downie:2011mm}
  E.~J.~Downie and H.~Fonvieille,
  Eur.\ Phys.\ J.\ ST {\bf 198} (2011) 287
  [arXiv:1106.0232 [nucl-ex]].

\bibitem{Huber:2015uza} 
  G.~M.~Huber and C.~Collicott,
  in: \textsc{12th Conference on the Intersections of
    Nuclear and Particle Physics (CIPANP 2015)}, Vail (USA), May 2015, eConf
  C15-05-19 (2015) [arXiv:1508.07919 [nucl-ex]].

\bibitem{Myers:2015aba} 
  L.~Myers, J.~Annand, J.~Brudvik, G.~Feldman, K.~Fissum, H.~Grie\3hammer, K.~Hansen and S.~Henshaw {\it et al.} [COMPTON@MAX-lab Collaboration],
  Phys.~Rev.~C \textbf{92} (2015)
  025203 [arXiv:1503.08094 [nucl-ex]].

\bibitem{Martel:2014pba} 
  P.~P.~Martel {\it et al.}  [A2 Collaboration],
  Phys.\ Rev.\ Lett.\  {\bf 114} (2015) 112501
  [arXiv:1408.1576 [nucl-ex]].

\bibitem{Chang:2015qxa} 
  E.~Chang {\it et al.} [NPLQCD Collaboration],
  Phys.\ Rev.\ D {\bf 92} (2015) 114502
  [arXiv:1506.05518 [hep-lat]].

\bibitem{Lujan:2014qga} 
  M.~Lujan, A.~Alexandru, W.~Freeman and F.~Lee,
  PoS LATTICE {\bf 2014} (2014) 153
  [arXiv:1411.0047 [hep-lat]].

\bibitem{Detmold:2010ts} 
  W.~Detmold, B.~C.~Tiburzi and A.~Walker-Loud,
  Phys.\ Rev.\ D {\bf 81} (2010) 054502
  [arXiv:1001.1131 [hep-lat]].

\bibitem{Primer:2013pva} 
  T.~Primer, W.~Kamleh, D.~Leinweber and M.~Burkardt,
  Phys.\ Rev.\ D {\bf 89} (2014) 034508
  [arXiv:1307.1509 [hep-lat]].

\bibitem{Hall:2013dva} 
  J.~M.~M.~Hall, D.~B.~Leinweber and R.~D.~Young,
  Phys.\ Rev.\ D {\bf 89} (2014) 054511
  [arXiv:1312.5781 [hep-lat]].

\bibitem{Engelhardt:2011qq} 
  M.~Engelhardt,
  PoS LATTICE {\bf 2011} (2011) 153
  [arXiv:1111.3686 [hep-lat]].

\bibitem{Engelhardt:2007ub} 
  M.~Engelhardt [LHPC Collaboration],
  Phys.\ Rev.\ D {\bf 76} (2007) 114502
  [arXiv:0706.3919 [hep-lat]].

\bibitem{Engelhardt:2010tm} 
  M.~Engelhardt,
  PoS LAT {\bf 2009} (2009) 128
  [arXiv:1001.5044 [hep-lat]].

\bibitem{Engelhardt:2015} M.~Engelhardt, J.~Saenz and R.~H\"ollwieser, private communication and forthcoming. 

\bibitem{Freeman:2014kka} 
  W.~Freeman, A.~Alexandru, M.~Lujan and F.~X.~Lee,
  Phys.\ Rev.\ D {\bf 90} (2014) 054507
  [arXiv:1407.2687 [hep-lat]].

\bibitem{Griesshammer:2014xla} 
  H.~W.~Grie\3hammer, A.~I.~L'vov, J.~A.~McGovern, V.~Pascalutsa, B.~Pasquini and D.~R.~Phillips,
  arXiv:1409.1512 [nucl-th].

\bibitem{WalkerLoud:2012bg}
  A.~Walker-Loud, C.~E.~Carlson and G.~A.~Miller,
  Phys.\ Rev.\ Lett.\  {\bf 108} (2012) 232301 
  [arXiv:1203.0254 [nucl-th]].

\bibitem{WalkerLoud:2012en} 
  A.~Walker-Loud, C.~E.~Carlson and G.~A.~Miller,
  PoS LATTICE {\bf 2012} (2012) 136
  [arXiv:1210.7777 [hep-lat]].

\bibitem{Erben:2014hza} 
  F.~B.~Erben, P.~E.~Shanahan, A.~W.~Thomas and R.~D.~Young,
  Phys.\ Rev.\ C {\bf 90} (2014) 065205
  [arXiv:1408.6628 [nucl-th]].

\bibitem{Thomas:2014dxa} 
  A.~W.~Thomas, X.~G.~Wang and R.~D.~Young,
  Phys.\ Rev.\ C {\bf 91} (2015) 015209
  [arXiv:1406.4579 [nucl-th]].

\bibitem{Gasser:2015dwa} 
  J.~Gasser, M.~Hoferichter, H.~Leutwyler and A.~Rusetsky,
  Eur.\ Phys.\ J.\ C {\bf 75} (2015) 375
  [arXiv:1506.06747 [hep-ph]].

\bibitem{Pachucki} K.~Pachucki, Phys.\ Rev.\ A \textbf{60} (1999) 3593.

\bibitem{Carlson:2011dz}
  C.~E.~Carlson and M.~Vanderhaeghen,
  arXiv:1109.3779 [physics.atom-ph].  

\bibitem{Pohl:2013yb} 
  R.~Pohl, R.~Gilman, G.~A.~Miller and K.~Pachucki,
  Ann.\ Rev.\ Nucl.\ Part.\ Sci.\  {\bf 63} (2013) 175
  [arXiv:1301.0905 [physics.atom-ph]].


\bibitem{Lensky:2014efa}
  V.~Lensky and J.~A.~McGovern,
  Phys.\ Rev.\ C {\bf 89} (2014)  032202
  [arXiv:1401.3320 [nucl-th]].

\bibitem{observables} H.~W.~Grie\3hammer, J.~A.~McGovern and D.~R.~Phillips,
  forthcoming.

\bibitem{Cacciari:2011ze} 
  M.~Cacciari and N.~Houdeau,
  JHEP {\bf 1109} (2011) 039
  [arXiv:1105.5152 [hep-ph]].

\bibitem{Furnstahl:2015rha} 
  R.~J.~Furnstahl, N.~Klco, D.~R.~Phillips and S.~Wesolowski,
  Phys.\ Rev.\ C {\bf 92} (2015) 024005
  [arXiv:1506.01343 [nucl-th]].

  \bibitem{Furnstahl:2014xsa} 
  R.~J.~Furnstahl, D.~R.~Phillips and S.~Wesolowski,
  J.\ Phys.\ G {\bf 42} (2015) 034028
  [arXiv:1407.0657 [nucl-th]].
  
\bibitem{Babusci:1998ww}
  D.~Babusci, G.~Giordano, A.~I.~L'vov, G.~Matone and A.~M.~Nathan,
  Phys.\ Rev.\  C {\bf 58} (1998) 1013
  [arXiv:hep-ph/9803347].

\bibitem{talkMAMI} H.~W.~Grie\3hammer, \emph{High-Accuracy Analysis of Compton
    Scattering in Chiral EFT; Status and Future}, invited seminar,
  \textsc{A2/Crystall-Ball Collaboration Meeting 2013}, Institut f\"ur
  Kernphysik, Johannes-Gutenberg-Universit\"at Mainz, Mainz (Germany), 4-5
  July 2013; talk, workshop on \textsc{Compton Scattering off Protons and
    Light Nuclei: Pinning Down the Nucleon Polarisabilities}, ECT*, Trento
  (Italy), 29 July-2 August 2013.

\bibitem{Rentmeester:1999vw} 
  M.~C.~M.~Rentmeester, R.~G.~E.~Timmermans, J.~L.~Friar and J.~J.~de Swart,
  Phys.\ Rev.\ Lett.\  {\bf 82} (1999) 4992
  [nucl-th/9901054].

\bibitem{bksm93}  
  V.~Bernard, N.~Kaiser, A.~Schmidt and U.-G.~Mei\3ner,
  Phys.\ Lett.\ B {\bf 319} (1993) 269
  [hep-ph/9309211].

\bibitem{VijayaKumar:2000pv}
  K.~B.~Vijaya Kumar, J.~A.~McGovern and M.~C.~Birse,
  Phys.\ Lett.\ B {\bf 479} (2000) 167
  [hep-ph/0002133].

\bibitem{Olmos:2001}
  V.~Olmos de Le\'on et al.,
  Eur.\ Phys.\ J.\ A {\bf 10 } (2001) 207.

\bibitem{Levchuk:1999zy} 
  M.~I.~Levchuk and A.~I.~L'vov,
  Nucl.\ Phys.\ A {\bf 674} (2000) 449
  [nucl-th/9909066].

\bibitem{Pascalutsa:2003zk}
  V.~Pascalutsa and D.~R.~Phillips,
  Phys.\ Rev.\ C {\bf 68} (2003) 055205
  [nucl-th/0305043].

\bibitem{hhk97}
  T.~R.~Hemmert, B.~R.~Holstein and J.~Kambor,
  Phys.\ Rev.\ D {\bf 55} (1997) 5598
  [hep-ph/9612374].

\bibitem{hhkk98}
  T.~R.~Hemmert, B.~R.~Holstein, J.~Kambor and G.~Knochlein,
  Phys.\ Rev.\ D {\bf 57} (1998) 5746
  [nucl-th/9709063].

\bibitem{Lensky:2015awa} 
  V.~Lensky, J.~McGovern and V.~Pascalutsa,
  Eur.\ Phys.\ J.\ C {\bf 75} (2015) 604
  [arXiv:1510.02794 [hep-ph]].

\bibitem{Ahrens:2001qt}
J.~Ahrens {\it et al.} [GDH and A2 Collaborations],
Phys.\ Rev.\ Lett.\ {\bf 87} (2001) 022003
[hep-ex/0105089].

\bibitem{Dutz} H.~Dutz, K.~Helbing, J.~Krimmer, T.~Speckner and
G.~Zeitler [GDH and A2 Collaborations], Phys.\ Rev.\ 
Lett.\ \textbf{91} (2003) 192001.

\bibitem{Camen} M.~Camen et al., Phys.~Rev.~C \textbf{65} (2002) 032202.

\bibitem{Holstein:1999uu}
  B.~R.~Holstein, D.~Drechsel, B.~Pasquini and M.~Vanderhaeghen,
  Phys.\ Rev.\  C {\bf 61} (2000) 034316
  [arXiv:hep-ph/9910427].

\bibitem{Pasquiniprivcomm} B.~Pasquini, private communication based on
  Ref.~\cite{Hildebrandt:2003fm}.

\bibitem{Pasquini:2010zr}
B.~Pasquini, P.~Pedroni and D.~Drechsel,
Phys.\ Lett.\ B {\bf 687} (2010) 160
[arXiv:1001.4230 [hep-ph]].

\bibitem{Schumacher:2005an}
  M.~Schumacher,
  Prog.\ Part.\ Nucl.\ Phys.\  {\bf 55} (2005) 567
  [arXiv:hep-ph/0501167].

\bibitem{Kossert:2002ws} 
  K.~Kossert, M.~Camen, F.~Wissmann, J.~Ahrens, J.~R.~M.~Annand, H.~J.~Arends, R.~Beck and G.~Caselotti {\it et al.},
  Eur.\ Phys.\ J.\ A {\bf 16} (2003) 259
  [nucl-ex/0210020].

\bibitem{Schindler:2008fh} 
  M.~R.~Schindler and D.~R.~Phillips,
  Annals Phys.\  {\bf 324} (2009) 682   [arXiv:0808.3643 [hep-ph]]; erratum, Annals Phys.\  {\bf 324} (2009) 2051.

\bibitem{Wesolowski:2015fqa} 
  S.~Wesolowski, N.~Klco, R.~J.~Furnstahl, D.~R.~Phillips and A.~Thapaliya,
  arXiv:1511.03618 [nucl-th].

  \bibitem{Epelbaum:2014}
    E.~Epelbaum, H.~Krebs and U.-G.~Mei\3ner, 
  Eur.\ Phys.\ J.\ A {\bf 51} (2015) 53
  [arXiv:1412.0142 [nucl-th]].

\bibitem{Epelbaum:2014sza} 
  E.~Epelbaum, H.~Krebs and U.-G.~Mei\3ner,
  Phys.\ Rev.\ Lett.\  {\bf 115} (2015) 122301
  [arXiv:1412.4623 [nucl-th]].
  
\bibitem{Binder:2015mbz} 
  S.~Binder {\it et al.},
  arXiv:1505.07218 [nucl-th].


\bibitem{Griesshammer:2000mi} 
  H.~W.~Grie\3hammer and G.~Rupak,
  Phys.\ Lett.\ B {\bf 529} (2002) 57
  [nucl-th/0012096].

\bibitem{dcompton-delta4} H.~W.~Grie\3hammer, J.~A.~McGovern and
  D.~R.~Phillips, forthcoming.

\bibitem{Procura:2006gq} 
  M.~Procura, B.~U.~Musch, T.~R.~Hemmert and W.~Weise,
  Phys.\ Rev.\ D {\bf 75} (2007) 014503
  [hep-lat/0610105].

\bibitem{McGovern:2006fm}
  J.~A.~McGovern and M.~C.~Birse,
  Phys.\ Rev.\ D {\bf 74} (2006) 097501
  [hep-lat/0608002].


\bibitem{Djukanovic:2006xc}
  D.~Djukanovic, J.~Gegelia and S.~Scherer,
  Eur.\ Phys.\ J.\ A {\bf 29} (2006) 337
  [hep-ph/0604164].


\bibitem{Schindler:2007dr}
  M.~R.~Schindler, D.~Djukanovic, J.~Gegelia and S.~Scherer,
  Nucl.\ Phys.\ A {\bf 803} (2008) 68
  [arXiv:0707.4296 [hep-ph]].

\bibitem{WalkerLoud:2008bp} 
  A.~Walker-Loud, H.-W.~Lin, D.~G.~Richards, R.~G.~Edwards, M.~Engelhardt, G.~T.~Fleming, P.~Hagler and B.~Musch {\it et al.},
  Phys.\ Rev.\ D {\bf 79} (2009) 054502
  [arXiv:0806.4549 [hep-lat]].

\bibitem{Walker-Loud:2013yua} 
  A.~Walker-Loud,
  PoS CD {\bf 12} (2013) 017
  [arXiv:1304.6341 [hep-lat]].
  
\bibitem{Bernard:2006te}
  V.~Bernard and U.~G.~Meissner,
  Phys.\ Lett.\ B {\bf 639} (2006) 278
  [hep-lat/0605010].


  \bibitem{Beane:2015}
  S.~Beane, talk at ``8th International Workshop on Chiral Dynamics: Theory
  and Experiment", Pisa, Italy, 29 June--3 July, 2015. 

\bibitem{Lensky:2009uv} 
  V.~Lensky and V.~Pascalutsa,
  Eur.\ Phys.\ J.\ C {\bf 65} (2010) 195
  [arXiv:0907.0451 [hep-ph]].


\bibitem{Beane:2014ora} 
  S.~R.~Beane {\it et al.},
  Phys.\ Rev.\ Lett.\  {\bf 113} (2014) 252001
  [arXiv:1409.3556 [hep-lat]].

\bibitem{Bedaque:2010hr} 
  P.~F.~Bedaque, T.~Luu and L.~Platter,
  Phys.\ Rev.\ C {\bf 83}, 045803 (2011)
  [arXiv:1012.3840 [nucl-th]].

\bibitem{Blum:2010ym} 
  T.~Blum, R.~Zhou, T.~Doi, M.~Hayakawa, T.~Izubuchi, S.~Uno and N.~Yamada,
  Phys.\ Rev.\ D {\bf 82} (2010) 094508
  [arXiv:1006.1311 [hep-lat]].

\bibitem{Detmold:2006vu} 
  W.~Detmold, B.~C.~Tiburzi and A.~Walker-Loud,
  Phys.\ Rev.\ D {\bf 73} (2006) 114505
  [hep-lat/0603026].

\bibitem{Tiburzi:2014zva} 
  B.~C.~Tiburzi,
  Phys.\ Rev.\ D {\bf 89} (2014) 074019
  [arXiv:1403.0878 [hep-lat]].

 \bibitem{Stump:2001gu}
  D.~Stump, J.~Pumplin, R.~Brock, D.~Casey, J.~Huston, J.~Kalk, H.~L.~Lai and W.~K.~Tung,
  Phys.\ Rev.\ D {\bf 65} (2001) 014012
  [hep-ph/0101051].

\bibitem{Lvov:1993fp}
  A.~I.~L'vov,
  Int.\ J.\ Mod.\ Phys.\  A {\bf 8} (1993) 5267.

\bibitem{Bawin:1996nz}
  M.~Bawin and S.~A.~Coon,
  Phys.\ Rev.\  C {\bf 55} (1997) 419
  [arXiv:nucl-th/9610028].

\bibitem{Lee:2014iha} 
  J.~W.~Lee and B.~C.~Tiburzi,
  Phys.\ Rev.\ D {\bf 90} (2014) 074036
  [arXiv:1407.8159 [hep-lat]].

\bibitem{engelprivcomm} M.~Engelhardt, private communication.

\bibitem{tiburziprivcomm} B.~C.~Tiburzi, private communication.

\bibitem{Leinweber:1999ig} 
  D.~B.~Leinweber, A.~W.~Thomas, K.~Tsushima and S.~V.~Wright,
  Phys.\ Rev.\ D {\bf 61} (2000) 074502
  [hep-lat/9906027].

\bibitem{Gellas:1998wx} 
  G.~C.~Gellas, T.~R.~Hemmert, C.~N.~Ktorides and G.~I.~Poulis,
  Phys.\ Rev.\ D {\bf 60}, 054022 (1999)
  doi:10.1103/PhysRevD.60.054022
  [hep-ph/9810426].

\bibitem{Pascalutsa:2006up} 
  V.~Pascalutsa, M.~Vanderhaeghen and S.~N.~Yang,
  Phys.\ Rept.\  {\bf 437}, 125 (2007)
  doi:10.1016/j.physrep.2006.09.006
  [hep-ph/0609004].

\bibitem{Yao:2016vbz} 
  D.~L.~Yao, D.~Siemens, V.~Bernard, E.~Epelbaum, A.~M.~Gasparyan, J.~Gegelia, H.~Krebs and U.~G.~Meißner,
  arXiv:1603.03638 [hep-ph].
\end{thebibliography}
\end{document}